\documentclass[10pt]{amsart}
\usepackage{latexsym}
\usepackage{amsmath}
\usepackage{amssymb}

\newtheoremstyle{st_bem} 
  {}                 
  {}                 
  {\normalfont}         
  {}                    
  {\bfseries}           
  {.}                   
  { }            
  {}                    

\newtheorem{theorem}{Theorem}[section]
\newtheorem{lemma}[theorem]{Lemma}

\newtheorem{corollary}[theorem]{Corollary}
\theoremstyle{st_bem}
\newtheorem{remark}[theorem]{Remark}

\def\C{\mathbb{C}}
\def\R{\mathbb{R}}
\def\N{\mathbb{N}}

\def\d{\mathrm{d}}

\newcommand{\Tr}{\operatorname{Tr}}

\newcommand{\supp}{\operatorname{supp}}
\newcommand{\linhul}{\operatorname{span}}

\newcommand{\core}{\operatorname{ker}}

\newcommand{\gradient}{\operatorname{grad}}
\newcommand{\divergence}{\operatorname{div}}
\def\a{a}
\def\A{A}
\def\BFDelta{\Delta}
\def\BFdelta{\delta}
\numberwithin{equation}{section}
\keywords{elasticity operator, trapped modes}
\subjclass{35P20}
\begin{document}
\title[Trapped modes in the elastic plate]{Trapped modes for the
  elastic plate with a perturbation of Young's modulus}
\author{Clemens F{\"o}rster}
\begin{abstract}
We consider a linear elastic plate
with stress-free boundary conditions in the limit of vanishing Poisson
coefficient. We prove that under a local change of Young's modulus infinitely
many eigenvalues arise in the essential spectrum which accumulate at a
positive threshold. We give estimates on the accumulation rate and on the
asymptotical behaviour of the eigenvalues.
\end{abstract}
\maketitle

\section{Introduction}
We consider a homogeneous and isotropic linear elastic medium on the domain $G = \R^2 \times J$ with $J =
(-\frac{\pi}{2},\frac{\pi}{2})$. For functions $u \in H^1(G;\C^3)$ we set 
\begin{equation*}
\epsilon(u) = \frac12 (\nabla u + (\nabla u)^T)
\end{equation*}
and\footnote{For two matrices $A,B \in \C^{3\times 3}$ we set $\langle
  A,B\rangle_{\C^{3\times 3}} = \sum_{i,j=1}^3 A_{ij}\overline{B}_{ji}$.} 
\begin{equation}\label{eq:a0}
  \a_0[u]=2\int_G \langle \epsilon(u),\epsilon(u)\rangle_{\C^{3\times 3}}\,\d x.
\end{equation}
This is the quadratic form of the elasticity operator
\begin{equation}\label{diffex}
  \A_0=-\left(\Delta+\gradient\divergence\right)
\end{equation}
in $L^2(G;\C^3)$ for zero Poisson coefficient with stress-free boundary conditions. Here
we have chosen a suitable set of units such that for Young's modulus $E$ it
holds $E = 2$.
Let $f \in L^\infty(\R^2;[0,1])$ be a function
of compact support which is extended to $G$ by $f(x) = f(x_1,x_2)$ for $x =
(x_1,x_2,x_3) \in G = \R^2\times J$. The
function $f$ describes a local perturbation consisting in an area of reduced Young's
modulus. For $\alpha \in (0,1)$ we consider 
the perturbed operator $\A_\alpha$ corresponding to the quadratic form
\begin{equation}\label{aalpha}
  \a_\alpha[u] = 2\int_G (1-\alpha f) \langle \epsilon(u),
  \epsilon(u)\rangle_{\C^{3\times 3}}\,\d x,\quad u \in
  H^1(G;\C^3).
\end{equation}
The perturbation gives rise to local oscillations of the plate which are
situated around the perturbation. These oscillations are called trapped
modes and correspond to eigenvalues of $\A_\alpha$ which are embedded into its
essential spectrum. We prove the existence of infinitely many eigenvalues
$\varkappa_k(\alpha), k \in \N$, of $\A_\alpha$ for $\alpha \in (0,1)$ which
accumulate to a certain threshold $\Lambda > 0$. Furthermore, we give two
sorts of estimates which characterise the asymptotical behaviour of the
eigenvalues. On the one hand we prove the small coupling asymptotics
\begin{equation}
  \varkappa_k(\alpha) = \Lambda - \alpha^2 (\Lambda\pi \lambda_k(K))^2 +
  o(\alpha^2) \quad\text{as}\quad \alpha \to 0.
\end{equation} 
Here $\lambda_k(K)$ are the eigenvalues of a compact model operator
$K$ specified in \eqref{eq:K}. On the other hand we prove the accumulation
rate asymptotics
\begin{equation}
  \ln (\Lambda-\varkappa_k(\alpha)) = -2k\ln k + o(k\ln k)\quad\text{as}\quad k \to \infty.
\end{equation}

The topic of this paper is related to a series of works on trapped modes in perturbed quantum,
acoustic and elastic waveguides, see among others
\cite{ELV,BGRS,RVW,GCA,EKK,ZPK,FW} and references
there\-in. While trapped modes for quantum and acoustic waveguides
correspond to eigenvalues of the scalar-valued Laplacian, the investigation of
trapped modes in elastic waveguides leads to the spectral analysis of the
elasticity operator which acts on vector-valued functions. Because of the more involved structure of this
operator, new and in some degree curious spectral effects can be observed.

These effects depend on the specific structure of the symbol which is associated with
the elasticity operator. In \cite{FW} this dependence was exploited for the
elasticity operator on the strip $\Gamma = \R \times J$. Performing a Fourier transform in the
unbounded direction on sees that the spectrum arises from an infinite series
of band functions depending on the Fourier parameter. Considering the operator
only for specific spatial and internal symmetry cases, the lowest band
function attains its minimum at two non-zero points $\xi = \pm \varkappa$ of
the Fourier coordinate $\xi$. This yields two trapped modes when a suitable
perturbation is applied. 

Within this article we consider the plate $G = \R^2 \times J$ which
is obtained by rotating the strip $\Gamma$ around the axis which is along the bounded
dimension. As we will see, the lowest band function in this case is obtained by rotating
the corresponding band function of the strip case around the
axis $\xi = 0$. Its minimum is now attained on the whole circle $\{ \xi \in
\R^2 : |\xi| = \varkappa\}$. This corresponds to infinitely many standing
waves with minimal kinetic energy. Therefore, indeed \emph{infinitely}
many eigenvalues arise, when a perturbation is applied. This kind of
``strongly degenerated'' spectral minimum was treated in
\cite{LSW} for rather general operators. We will make use of the concepts
which were developed there. But in contrast to the applications presented in \cite{LSW},
we have to deal with super-polynomial accumulation rates of the
eigenvalues. Such accumulation rates were recently observed in different
situations, see for example \cite{MR,RW,FP}. In general, they rest on the combination of
compactly supported perturbations applied to strongly degenerated operators. 

The structure of the paper is as follows. In Section \ref{sec:state} we state
the problem in detail. In Section \ref{sec:aux1} we describe the
spatial and internal symmetries of the problem which will help us to distinguish the
embedded eigenvalues from the surrounding essential spectrum. We give a
description of the spectral minimum mentioned above which is mainly based on \cite{FW}. 

In Section \ref{sec:exist} we prove
the existence of infinitely many eigenvalues. Further auxiliary material is
provided in Section \ref{sec:aux2}. We introduce
general classes of operators which allow us to handle
super-polynomial eigenvalue asymptotics beyond the well-known
$\Sigma_p$-classes for compact operators. Then we state a special type
of Birman-Schwinger principle appropriate to our problem. Finally we
introduce the method of \cite{LSW} and describe the behaviour of the
Birman-Schwinger operator along the spectral minimum of the unperturbed operator. 

The main results about the
asymptotical behaviour of the embedded eigenvalues are stated in
Section \ref{sec:main}, followed by the main proofs in Section \ref{sec:mainproof}. The idea 
of the proofs is to obtain properties from the spectrum of
the unbounded elasticity operator by investigating the compact Birman-Schwinger operator. This is done by
extracting that part of the Birman-Schwinger operator which is responsible for the leading
term in the eigenvalue asymptotics, calculating its properties and estimating the remainder
terms. The latter two problems are treated separately in Sections \ref{sec:K}
and \ref{sec:remainder}.

\subsection{Acknowledgements}
The author thanks T. Weidl for introducing him to this interesting
topic, for his guidance and for many valuable hints and discussions. 
The author is also grateful to A. Laptev for useful discussions.
The kind hospitality of the KTH is gratefully acknowledged. 
This work was financially supported by the
Cusanuswerk, the DAAD-STINT, PPP-programme and the DFG, grant We 1964/2-1.
\subsection{Notations}
Throughout this article, $n(s,T)$ denotes
the number of singular values of a compact operator $T$ above $s>0$ including multiplicities. For
general selfadjoint operators $T$, $n_\pm(s,T)$ denotes the number of
eigenvalues above and below $s\in \R$, respectively.
\section{Statement of the problem}\label{sec:state}
We denote the stress tensor by $\sigma(u) = \lambda(\Tr \epsilon(u))\mathrm{I} +2\mu \epsilon(u)$ where
$\mu = \frac{E}{2(1+\nu)}$ and $\lambda = \frac{E\nu}{(1+\nu)(1-2\nu)}$ are
the Lam{\'e} constants, $\nu$ is the Poisson ratio and $E$ is Young's
modulus. Then the quadratic form corresponding to the operator of linear
elasticity subject to stress-free boundary conditions is given by
\begin{equation}\label{a}
\a_0[u]=\int_G \langle \sigma(u) , \epsilon(u)\rangle_{\C^{3\times 3}}\,\d x,
\end{equation}
which is well-defined for all functions $u\in H^1(G;\C^3)$. 

In this paper we stress on the special case of zero Poisson coefficient.
Choosing a suitable set of units such that $E = 2$, \eqref{a} turns into
\begin{equation}\label{aa}
\a_0[u]=2\int_G\langle \epsilon(u) , \epsilon(u)\rangle_{\C^{3\times 3}}\,\d x.
\end{equation}
This form is associated with the
positive self-adjoint operator
\begin{equation}
  \A_0 = -(\Delta + \gradient \divergence)
\end{equation}
on the domain
\begin{equation*}
D(\A_0)=\left\{
u\in H^2(G;\C^3): \partial_j u_3+\partial_3u_j = 0,\; j=1,2,3
\quad\text{on}\quad \R^2\times \{\pm {\textstyle \frac{\pi}{2}}\}
\right\}.
\end{equation*}
The inequality
\begin{equation}\label{obv}
\a_0[u]\leq 2 \|u\|^2_{H^1(G;\C^3)} ,
\qquad u\in H^1(G;\C^3)
\end{equation}
is obvious. Moreover, the reverse estimate
\begin{equation}\label{korn}
\a_0[u]+\|u\|^2\geq
c(G) \|u\|^2_{H^1(G;\C^3)},
\qquad  u\in H^1(G;\C^3),\quad c(G)>0
\end{equation}
holds, where $\|\cdot\|$ denotes the norm in $L^2(G;\C^3)$. This is the 
well-known Korn inequality \cite{Gob}.
Hence, the class of functions $u\in  L^2(G;\C^3)$, for which the integral \eqref{aa} is well-defined and
finite, coincides with $H^1(G;\C^3)$. Therefore, the form $\a_0$ is closed on
the domain $D[\a_0] = H^1(G;\C^3)$.

For $f \in L^\infty(\R^2;[0,1])$ with compact support, extended to G by $f(x)
= f(x_1,x_2)$ for $x \in G=\R^2\times J$, we define the form 
\begin{equation}\label{v}
  v[u] := 2\int_G f \langle \epsilon(u) , \epsilon(u)\rangle_{\C^{3\times 3}}\,\d x,\qquad u \in D[\a_0].
\end{equation}
Considering the form 
\begin{equation}
\a_\alpha[u] := \a_0[u] - \alpha v[u]
\end{equation}
for $\alpha \in (0,1)$, which corresponds to \eqref{aalpha}, we see that this form is also closed on
the domain $D[\a_\alpha] = D[\a_0] = H^1(G;\C^3)$. Therefore it induces 
a positive self-adjoint operator $\A_\alpha$ in $L^2(G;\C^3)$.

The spectrum of the operator $\A_0$ is purely absolutely continuous and
coincides with $[0,\infty)$. It is well-known \cite{B},
that a local change of the boundary conditions or a local change of the
quadratic form will not change the essential spectrum. Therefore the 
essential part of the
spectrum of $A_\alpha$ fills the non-negative semi-axis, too.
In this paper we shall discuss the existence of positive eigenvalues of the operator $A_\alpha$
which are embedded into its continuous spectrum.
\section{Auxiliary material I}\label{sec:aux1}
\subsection{Spatial and internal symmetries}\label{sec:sym}
In analogy to \cite{FW} we define the subspaces ${H}_j$, $j=1,2$ of $H := L^2(G;\C^3)$ as
\begin{equation*}
  \begin{split}
    H_j := \big\{u\in{H}:\, &u_{l}(\cdot,-x_3)=(-1)^{j-1}u_{l}(\cdot,x_3),\;l=1,2,\\
    &u_3(\cdot,-x_3) = (-1)^j u_3(\cdot,x_3) \big\}.
  \end{split}
\end{equation*}
Then ${H}={H}_1\oplus {H}_2$. Further let
\begin{equation*}
{H}_{3}:=\{u\in H_1:\ u=(u_1(x_1,x_2),u_2(x_1,x_2),0)\}\ .
\end{equation*}
It forms a subspace in ${H}_1$. The orthogonal complement ${H}_{4}$ to
${H}_{3}$ in ${H}_1$
consists of all functions $w\in{H}_1$ for which
\begin{equation*}
\int_J w_i(\cdot,x_3)\,\d x_3 =0\quad\text{in}\quad
L^2(\R^2;\C)\quad\text{for}\quad i=1,2.
\end{equation*}
Let $P_j$ be the orthogonal projections onto ${H}_j$, $j=1,\dots ,4$.
Then $P_jP_1=P_1P_j=P_j$ for $j=3,4$.
A simple calculation
shows, that
\begin{equation*}
  D[\a_\alpha^{(j)}]:=P_j D[\a_\alpha]\subset
  D[\a_\alpha]\ ,\quad j=1,\dots ,4
\end{equation*}
and
\begin{equation*}
  \a_\alpha[u,w]=0\quad\text{for all}
  \quad u\in D[\a_\alpha^{(l)}],
  \ w\in D[\a_\alpha^{(j)}]
\end{equation*}
if $l,j=2,3,4$ and $l\neq j$. Hence,
these subspaces are reducing for the operator $\A_\alpha$ and
\begin{equation}\label{finaldec}
\A_\alpha=\A^{(3)}_\alpha\oplus \A^{(4)}_\alpha\oplus \A^{(2)}_\alpha
\quad\text{on}\quad
H=H_3\oplus H_4\oplus H_2\ ,
\end{equation}
where the operators $\A_\alpha^{(j)}$ are the restrictions of
$\A_\alpha$ to $D(\A_\alpha^{(j)})=D(\A_\alpha)\cap H_j$
and correspond to the closed forms
$\a_\alpha^{(j)}$, given by
the differential expression \eqref{aalpha} on
$D[\a_\alpha^{(j)}]$, $j=2,3,4$.
Put
\begin{equation}\label{34}
\A^{(1)}_\alpha=\A^{(3)}_\alpha\oplus \A^{(4)}_\alpha
\quad\mbox{on}\quad H_1=H_3\oplus H_4\ ,
\end{equation}
being the restriction of
$\A_\alpha$ to $D(\A_\alpha)\cap H_1$.
Then it holds
\begin{equation}\label{H12}
\A_\alpha=\A^{(1)}_\alpha\oplus \A^{(2)}_\alpha\quad
\text{on}\quad H=H_1\oplus H_2\ .
\end{equation}
Decomposition \eqref{H12} reflects the spatial symmetry of the operator $\A_\alpha$,
while \eqref{34} exploits the specific internal structure of
$\A_\alpha$. These symmetries depend on the above-mentioned restriction
to perturbation functions $f$ which are constant in $x_3$-direction.
We also note that \eqref{34} fails for elasticity operators with non-zero Poisson
coefficients. 

The hierarchy of symmetry spaces $H_j$ for the vector-valued
displacement functions carries over to the matrix-valued stress
functions. We set
\begin{equation*}
  H^\times = \{w \in L^2(G;\C^{3\times 3}) : w = w^T\}
\end{equation*}
and define the subspaces $H_j^\times$ of $H^\times$ by
\begin{equation*}
  \begin{split}
    H_1^\times &:= \{w \in H^\times :\,w_{ij}(\cdot,-x_3) = \delta
    w_{ij}(\cdot,x_3), \text{where}\\
    &\qquad\delta = 1\;\text{for}\; i,j \in \{1,2\}\;\text{or}\;
    i=j=3\;\text{and}\;\delta = -1\;\text{otherwise}\},\\
    H_4^\times &:= \{w \in H_1^\times :\, \int_J w_{ij}(\cdot,x_3)\,\d x_3 =
    0\;\text{in}\; L^2(\R^2;\C)\;\text{for}\; i,j \in \{1,2\}\},\\
    H_2^\times &:= H^\times \ominus H_1^\times,\qquad H_3^\times :=
    H_1^\times \ominus H_4^\times.
  \end{split}
\end{equation*}
Then 
\begin{equation*}
H^\times = H_3^\times \oplus H_4^\times \oplus H_2^\times
\end{equation*}
and
\begin{equation*}
  \epsilon(u) \in H_j^\times\quad\text{for}\quad u \in D[a_0] \cap H_j.
\end{equation*}

For functions $u \in D(Q) := D[\a^{(4)}_0]$ we define the operator
$Qu = \sqrt{2} \epsilon(u)$. It holds that $\A^{(4)}_0 = Q^\ast Q$.
In Section \ref{sec:bsprinc} we will need the operator
\begin{equation}\label{eq:U}
  U := Q\left(\A^{(4)}_0\right)^{-\frac12},\qquad U : H_4 \rightarrow H_4^\times
\end{equation}
which is an isometry between $H_4$ and the range of $U$,
\begin{equation*}
  R(U) = H_4^\times \ominus (\core Q^\ast).
\end{equation*}
This follows from
\begin{equation*}
  \|U v\|_\times^2 = \|Q(\A^{(4)}_0)^{-\frac12} v\|_\times^2 =
  \|\sqrt{2}\epsilon((\A^{(4)}_0)^{-\frac12} v)\|_\times^2 =
  \a^{(4)}_0[(\A^{(4)}_0)^{-\frac12} v] = \|v\|^2
\end{equation*}
for all $v \in H_4$ where $\|\cdot\|_\times$ denotes the norm in
$L^2(G;\C^{3\times 3})$. 

\subsection{Separation of variables for $\A_0$} 
Applying the unitary Fourier transform $\Phi$ in $(x_1,x_2)$-direction
and its inverse $\Phi^*$,
one finds that $\Phi \A_0\Phi^*$ permits the orthogonal
decomposition
\begin{equation*}
\Phi \A_0\Phi^*=
\int^{\oplus}_{\R^2} \A(\xi)\,\d\xi
\qquad\text{on}\qquad {H}=\int^\oplus_{\R^2} h\,\d\xi,\quad h=L^2(J;{\C}^3).
\end{equation*}
The self-adjoint operators $\A(\xi)$
are given by the differential expressions
\begin{equation}\label{diffxi}
  \A(\xi) = \begin{pmatrix} -\partial_3^2 + |\xi|^2+\xi_1^2 & \xi_1\xi_2 & -i\xi_1\partial_3\\
    \xi_1\xi_2 & -\partial_3^2 +|\xi|^2 +\xi_2^2& -i\xi_2\partial_3 \\ -i\xi_1\partial_3 & -i\xi_2\partial_3 & -2\partial_3^2 + |\xi|^2
  \end{pmatrix} 
\end{equation}
on the domains
\begin{equation}\label{dom}
  \begin{split}
    D(A(\xi))= \{u\in H^2(J;\C^3): \partial_3 u_3|_{t=\pm \pi/2} = &(\partial_3 u_1 + i\xi_1 u_3)|_{t=\pm \pi/2} = 0,\\
      &(\partial_3 u_2 + i\xi_2 u_3)|_{t=\pm \pi/2} = 0 \}.
  \end{split}
\end{equation}

The symmetry \eqref{finaldec} extends to the
operators $A(\xi)$. Indeed, put
\begin{align*}
  h_j &:= \{u \in h : u_{1/2}(t) =
  (-1)^{j-1}u_{1/2}(-t),u_3(t) = (-1)^j u_3(-t)\},\; j=1,2,\\
  h_3 &:= \linhul\{ (1, 0,0)^T, (0,1,0)^T\},\quad  h_4 := \{u \in h_1 : \int_J u_i(t)\,\mathrm{d}t = 0,\; i=1,2\}.
\end{align*}
Then we have
\begin{equation}\label{adeco}
  H_j=\int_{\R^2}^\oplus h_j\,\d\xi\qquad\text{and}\qquad
  \Phi \A^{(j)}_0\Phi^* = \int_{\R^2}^\oplus \A^{(j)}(\xi)\,\d\xi,
  \qquad j=1,\dots,4,
\end{equation}
where the operators $\A^{(j)}(\xi)$ are the restrictions of
$\A(\xi)$ to $D(\A^{(j)}(\xi))=D(\A(\xi))\cap h_j$. Moreover, it holds
\begin{equation}\label{xideco}
  \begin{split}
    \A(\xi)=\A^{(1)}(\xi)\oplus \A^{(2)}(\xi) \quad &\text{on} \quad
    h=h_{1}\oplus h_2\ ,\\ 
    \A(\xi)= \A^{(3)}(\xi)\oplus \A^{(4)}(\xi)\oplus \A^{(2)}(\xi) \quad
    &\text{on} \quad h=h_{3}\oplus h_{4}\oplus h_2\ .
  \end{split}
\end{equation}
The operators $\A^{(j)}(\xi)$
correspond to quadratic forms $\a^{(j)}(\xi)$
being closed on the domains $D[\a^{(j)}(\xi)]=
H^1(J;\C^3)\cap h_{j}$, $j=1,\dots,4$.
\subsection{The spectral analysis of the operator $A_0^{(4)}$}\label{sec:specanal}
During this paper the spectral decomposition of the operator $\A_0^{(4)}$
shall be of particular interest. Because of the decomposition
\eqref{adeco} we have in fact to carry out the spectral analysis of the
operators $\A^{(4)}(\xi)$.
Being the restrictions of the non-negative second order
Sturm-Liouville systems \eqref{diffxi} to $D(A(\xi))\cap h_{4}$,
the operators $\A^{(4)}(\xi)$ have
a non-negative discrete spectrum, which accumulates to infinity only.
The eigenvalues of $A^{(4)}(\xi)$ are invariant under rotations of $\xi$. If
$\lambda$ is an eigenvalue of $A^{(4)}(\xi)$ with eigenfunction $u(\xi) =
(u_1(\xi),u_2(\xi),u_3(\xi))^T$, then, for any unitary matrix $M \in \R^{2\times 2}$,
$\lambda$ is also an eigenvalue of $A^{(4)}(M\xi)$ with the eigenfunction
$u(M\xi)$ fulfilling
\begin{equation*}
  \binom{u_1(M\xi)}{u_2(M\xi)} = M\binom{u_1(\xi)}{u_2(\xi)},\qquad u_3(M\xi)
  = u_3(\xi).
\end{equation*}
Therefore, we can restrict ourselves to the special case $\xi =
\binom{0}{r}, r \geq 0$ where
\begin{equation*}
  \A^{(4)}(\xi) = \A^{(4)}(r) = \begin{pmatrix} -\partial_3^2 + r^2 & 0 & 0\\
    0 & -\partial_3^2 +2r^2 & -ir\partial_3 \\ 0 & -ir\partial_3 & -2\partial_3^2 + r^2
  \end{pmatrix}.
\end{equation*}
The operator $\A^{(4)}(r)$ can be written as $\A^{(4)}(r) = \hat{A}^{(4)}(r) \oplus
\check{A}^{(4)}(r)$ where the latter two operators are the restrictions of
$A^{(4)}(r)$ to 
\begin{equation*}
  \begin{split}
    &D(\hat{A}^{(4)}(r)) = \{u \in D(\A^{(4)}(r)):u_1 = 0\},\\
    &D(\check{A}^{(4)}(r)) = \{u \in D(\A^{(4)}(r)):u_2 = u_3 = 0\}.
  \end{split}
\end{equation*}
For the eigenvalues $\check{\lambda}_k(r)$ of $\check{A}^{(4)}(r)$ it holds
\begin{equation}\label{eq:evcheckA}
  \check{\lambda}_k(r) = r^2+4k^2,\quad k \in \N.
\end{equation}
The eigenvalues $\hat{\lambda}_k(r)$ of the operator $\hat{A}^{(4)}(r)$
however exhibit a nontrivial structure which was analysed in \cite{FW}. The eigenvalues are simple for any
fixed $r$ and the functions $r \mapsto \hat{\lambda}_k(r)$ are real
analytic. The spectral minimum
\begin{equation}
  \Lambda = \inf \bigcup_{r \geq 0} \sigma(\hat{A}^{(4)}(r)) = \inf_{r \geq 0}
  \hat{\lambda}_1(r)
\end{equation}
is achieved for exactly one non-trivial value $r = \varkappa > 0$. Moreover 
it holds 
\begin{equation}
  \hat{\lambda}_1(\varkappa+\varepsilon) = \Lambda + q^2\varepsilon^2+
  O(\varepsilon^3)\quad\text{as}\quad \varepsilon \to 0
\end{equation}
for a certain $q > 0$.
Applying the function $u(t) = (0,1-\frac{\pi}{2}\cos(t), i\frac{\pi}{2}\sin(t))^T$ to the quadratic
form $\hat{a}^{(4)}(r)$ associated with $\hat{A}^{(4)}(r)$ yields for $r = 1$ 
\begin{equation*}
  \frac{\hat{a}^{(4)}(1)[u]}{\|u\|^2_{L^2(J;\C^3)}} = 2
\end{equation*}
and therefore $\Lambda \leq 2$. Comparing this with \eqref{eq:evcheckA} we
achieve for the ground state $\lambda_1(r)$ of $\A^{(4)}(r)$
\begin{equation*}
  \Lambda = \inf \bigcup_{r \geq 0} \sigma(\A^{(4)}(r)) = \inf_{r \geq 0} \lambda_1(r)
\end{equation*}
where the infimum is attained exactly at $r = \varkappa$. As stated in
\cite{FW}, it holds
\begin{equation}\label{eq:lambda1}
  \lambda_1(\varkappa+\varepsilon) = \Lambda + q^2\varepsilon^2+
  O(\varepsilon^3)\quad\text{as}\quad \varepsilon \to 0
\end{equation}
for some $q > 0$. A numerical evaluation gives
\begin{equation}
  \begin{split}
    {\varkappa} &= 0.632138\pm 10^{-6}\ ,\\
    \Lambda &= 1.887837 \pm 10^{-6}\ ,\\
    q&= 0.849748 \pm 10^{-6}.
\end{split}
\end{equation}
Also from \cite{FW} we obtain that the eigenfunction corresponding to
$\lambda_1(r)$ can be given by $\psi_1
= \tilde{\psi}_1/\|\tilde{\psi}_1\|_{L^2(J;\C^3)}$ where
\begin{equation*}
  \tilde{\psi}_1(r,x_3) := \big(0,ir\, d_1(r,x_3), d_2(r,x_3)\big)^T
\end{equation*}
and
\begin{equation}
\begin{split}
   d_1(r,t) &:= {\textstyle r\beta\cos\left(\frac{\pi}{2}\beta\right)\cos(\gamma t) +
    \frac{\gamma^2\beta}{r}\cos\left(\frac{\pi}{2}\gamma\right)\cos(\beta t)},\\
   d_2(r,t) &:=  {\textstyle -r\beta\gamma\cos\left(\frac{\pi}{2}\beta\right)\sin(\gamma t) +
    r\gamma^2\cos\left(\frac{\pi}{2}\gamma\right)\sin(\beta t)}
\end{split}
\end{equation}
with $\beta = \sqrt{\lambda_1(r)-r^2}$ and $\gamma =
\sqrt{\lambda_1(r)/2-r^2}$. Note that this is only valid for $\gamma
\neq 0$ which is fulfilled in a neighbourhood of $r=\varkappa$. 

For general $\xi \in \R^2$ in a neighbourhood of $|\xi| = \varkappa$ we denote the eigenfunction
corresponding to $\lambda_1(\xi)$ again by $\psi_1
= \tilde{\psi}_1/\|\tilde{\psi}_1\|_{L^2(J;\C^3)}$ where now
\begin{equation}\label{eq:psi1}
  \tilde{\psi}_1(\xi,x_3) = \big(i\xi_1\,d_1(|\xi|,x_3),i\xi_2\, d_1(|\xi|,x_3), d_2(|\xi|,x_3)\big)^T.
\end{equation}
Moreover, the spectral minimum of $A^{(4)}(\xi)$ is attained for $|\xi| =
\varkappa$ where $\lambda_1(\xi) = \Lambda$. For such $\xi$ the functions 
\begin{equation}\label{eq:wxi}
  w_\xi(x) = \psi_1(\xi,x_3)\mathrm{e}^{i\xi\cdot \binom{x_1}{x_2}}
\end{equation}
fulfil
\begin{equation}
  -(\Delta + \gradient \divergence)w_\xi(x) = \Lambda w_\xi
  (x)\quad\text{for}\quad x \in G.
\end{equation}
\section{Existence of eigenvalues}\label{sec:exist}
\begin{theorem}\label{th:exist}
  If $f \not\equiv 0$ in $L^2$-sense, then $\A^{(4)}_\alpha$ has infinitely many
  eigenvalues in $[0,\Lambda)$ for all $\alpha \in (0,1)$.
\end{theorem}
The idea of the proof is to construct suitable test spaces from the functions
$w_\xi$ in \eqref{eq:wxi} such that the quadratic form of $A_\alpha^{(4)}$ is strictly lower than
$\Lambda$ for all normed functions of these
spaces. Then the proof follows from an application of the variational principle.

\subsection{Proof of Theorem \ref{th:exist}}
First we define suitable cut-off functions for $w_\xi$. Let $\tilde{\zeta} \in
\mathrm{C}^\infty(\R)$ with $\tilde{\zeta}(t) = 1$ for $t < 1$,
$\tilde{\zeta}(t) = 0$ for $t > 2$ and $0 \leq \tilde{\zeta}(t) \leq 1$ for
$t \in (1,2)$ and set
\begin{equation*}
  \zeta_\varepsilon(x) := \tilde{\zeta}(\varepsilon \ln |x|)\quad\text{for}\quad x \in
  \R^2,\, \varepsilon > 0.
\end{equation*}
It holds $\zeta_\varepsilon \in \mathrm{C}_0^\infty(\R^2)$ and 
\begin{equation}\label{eq:zetatozero}
\int_{\R^2} |\nabla \zeta_\varepsilon|^2\,\d x \to
0\quad\text{for}\quad \varepsilon \to 0.
\end{equation}

Next, for arbitrary $m\in \N$ we choose $\xi^k \in \R^2, k=1,\dots,m$
with $|\xi^k| = \varkappa$ such that $\xi_1^k \neq \xi_1^l$ for $k \neq
l$. We set
\begin{equation*}
  u_k^{(\varepsilon)}(x) =
  \zeta_\varepsilon(x_1,x_2)w_{\xi^k}(x)\quad\text{for}\quad x
  \in G
\end{equation*}
and define the test space
\begin{equation*}
  \mathcal{E}_m^{(\varepsilon)} =
  \linhul \{u_k^{(\varepsilon)} : k = 1,\dots,m\}.
\end{equation*}
As we show below,
\begin{equation}\label{eq:dimm}
  \dim \mathcal{E}_m^{(\varepsilon)} = m\quad\text{for all}\quad
  \varepsilon > 0.
\end{equation}
Therefore every element of $\mathcal{E}_m^{(\varepsilon)}$ can be
represented in a unique way as 
\begin{equation*}
  u_\eta = \sum_{k=1}^m \eta_k u_k^{(\varepsilon)}\quad\text{for}\quad
  \eta \in \C^m.
\end{equation*}

We recall that $a_\alpha[u] = a_0[u] - \alpha v[u]$ for $u\in D[a_0]$ where the quadratic form
$v$ was defined in \eqref{v}. As we will prove below it holds
\begin{equation}\label{eq:infv}
  \inf_{0 \neq u_\eta \in \mathcal{E}_m^{(\varepsilon)}}
  \frac{v[u_\eta]}{|\eta|^2} \geq c_0
\end{equation}
for some $c_0 > 0$ independent of sufficiently small $\varepsilon > 0$ and 
\begin{equation}\label{eq:supa}
  \sup_{0 \neq u_\eta \in \mathcal{E}_m^{(\varepsilon)}}
  \frac{a_0^{(4)}[u_\eta] - \Lambda\|u_\eta\|^2}{|\eta|^2} \to
  0\quad\text{as}\quad \varepsilon \to 0.
\end{equation}
Then for sufficiently small $\varepsilon > 0$ there exists
$\delta(\varepsilon) \in (0,\alpha c_0)$ such that 
\begin{equation*}
  \begin{split}
    a_\alpha^{(4)}[u_\eta] &= a_0^{(4)}[u_\eta] - \alpha v[u_\eta] =
    a_0^{(4)}[u_\eta] - \Lambda \|u_\eta\|^2 - \alpha v[u_\eta] +
    \Lambda\|u_\eta\|^2\\
    &< \delta(\varepsilon)|\eta|^2 - \alpha c_0|\eta|^2 + \Lambda
    \|u_\eta\|^2\\
    &< \Lambda \|u_\eta\|^2
  \end{split}
\end{equation*}
for all $u_\eta \in \mathcal{E}_m^{(\varepsilon)}$. From Glazman's
Lemma \cite[10.2.2]{BS} it follows the existence of at least $m$ eigenvalues (including
multiplicities) below $\Lambda$. Because $m$ was arbitrary, this proves the theorem.

\subsection{Proof of \eqref{eq:dimm}}\label{sec:dimm}
Choose an interval $I\subset \R$ and $x_2 \in \R$ such
that $\zeta_\varepsilon|_{I \times \{x_2\}} = 1$. Recall the
definition of $w_\xi$ in \eqref{eq:wxi} and of $\psi_1$ in
\eqref{eq:psi1}. Choose $x_3 \in J$ such that $d_2(\varkappa,x_3) \neq 0$. Then for arbitrary
$\eta \in \C^m$ the condition
\begin{equation*}
  \sum_{k=1}^m \eta_k u_k^{(\varepsilon)}(x) = 0, \quad x \in
    \R^2\times J
\end{equation*}
evaluated for $x \in I \times \{x_2\} \times \{x_3\}$ implies 
\begin{equation*}
  \sum_{k=1}^m \eta_k \mathrm{e}^{i\xi_1^k x_1} = 0,\quad x_1 \in I.
\end{equation*}
Because $\xi_1^k \neq \xi_1^l$ for $k \neq l$ the functions $x_1
\mapsto \mathrm{e}^{i\xi_k x_1}, k=1,\dots,m$ are linearly independent. Hence,
$\eta = 0$. This proves the linear independence of the functions
$u_k^{(\varepsilon)}$.

\subsection{Proof of \eqref{eq:infv}}
Because $f \neq 0$ in $L^2$-sense there exists
$\Omega \subset \R^2$ open, nonempty and $c > 0$ such that for $\chi_\Omega$ being
the characteristic function of $\Omega$ it holds $c\chi_\Omega(x) \leq
f(x)$ for almost every $x\in \R^2$.
Hence,
\begin{equation*}
  \inf_{|\eta| = 1} v[u_\eta] = \inf_{|\eta| = 1}\, 2\int_G f |\epsilon(u_\eta)|^2\,\d x \geq \inf_{|\eta|
    = 1}\, 2c \int_{\Omega\times J} |\epsilon(u_\eta)|^2\,\d x =: c_0.
\end{equation*}
We consider only sufficiently small $\varepsilon > 0$ such that
$\zeta_\varepsilon|_\Omega = 1$. Then $c_0$ does not depend on $\varepsilon$.
If $c_0 = 0$ there must be some function $v_\eta \in
\mathcal{E}_m^{(\varepsilon)}, |\eta| = 1$ such that 
$\int_{\Omega\times J} |\epsilon(v_\eta)|^2\,\d x = 0$. Particularly
this means
\begin{equation*}
  0 = \partial_3 u_{\eta,3}(x) = \sum_{k=1}^m \eta_k \partial_3 u_{k,3}^{(\varepsilon)}(x)\quad \text{for}\quad x \in \Omega
  \times J.
\end{equation*}
In analogy to Subsection \ref{sec:dimm} we obtain $\eta = 0$ which
is obviously contrary to the requirement $|\eta| = 1$. This proves $c_0 > 0$.
  
\subsection{Proof of \eqref{eq:supa}}
Note that $\| u \|_a = \left( a_0^{(4)} [u] - \Lambda
\|u\|^2\right)^{\frac12}$ for $u \in D[a_0^{(4)}]$ defines a norm 
in $D[a_0^{(4)}]$. Therefore it holds that
\begin{equation}\label{eq:sumueta}
    \| u_\eta\|_a^2 \leq \left(\sum_{k=1}^m |\eta_k|\,
      \|u_k^{(\varepsilon)}\|_a\right)^2 \leq |\eta|^2 \sum_{k=1}^m \|
    u_k^{(\varepsilon)}\|_a^2.
\end{equation}
A direct calculation using the definitions of $w_\xi$ in \eqref{eq:wxi} and 
$\psi_1$ in \eqref{eq:psi1} yields
\begin{equation*}
  \begin{split}
    \|u_k^{(\varepsilon)}\|^2_a &= \a_0^{(4)}[u_k^{(\varepsilon)}]-
    \Lambda\|u_k^{(\varepsilon)}\|^2\\
    &=\int_G (2|\psi_{1,1}|^2+|\psi_{1,3}|^2)|\partial_1 \zeta_\varepsilon|^2 + 
    (2|\psi_{1,2}|^2+|\psi_{1,3}|^2)|\partial_2 \zeta_\varepsilon|^2+\\
    &\phantom{=\int_G } +|\psi_{1,2}\,\partial_1
    \zeta_\varepsilon + \psi_{1,1}\,\partial_2\zeta_\varepsilon|^2\,\d x.
  \end{split}
\end{equation*}
Hence,
\begin{equation*}
\|u_k^{(\varepsilon)}\|^2_a\leq C \int_{\R^2}|\nabla \zeta_\varepsilon|^2\,\d (x_1,x_2)
\end{equation*}
for $C>0$ independent of $\varepsilon$. The result follows now from
\eqref{eq:zetatozero} and \eqref{eq:sumueta}.
\section{Auxiliary material II}\label{sec:aux2}
\subsection{The classes $\Sigma_\pi$ and $\mathcal{S}^\pi$}\label{sec:Sp}
In order to describe the exponential accumulation rate of the eigenvalues of
$\A_\alpha^{(4)}$ at $\Lambda$ we need to generalise the well-known weak
Neumann-Schatten classes $\Sigma_p$, see \cite[11.6]{BS}, and the classes $\mathcal{S}^p$
introduced by Safronov \cite{Sa,LSW}. They are both designed to measure polynomial
accumulation rates only. We develop these generalisations only as far
as it is needed here. Particularly we abandon to postulate conditions
ensuring basic properties like linearity or ideal structure for the generalised
classes. For a systematic treatment of such extensions see for example \cite{GK,W}.

Let $\pi : \R_+ \to \R_+$ be a function with $\lim_{s\to 0} \pi(s)
= 0$. Then for suitable compact operators $T$ we can define the functionals
\begin{equation*}
  \Delta_\pi(T) = \limsup_{s \to 0} \pi(s) n(s,T),\quad
  \delta_\pi(T) = \liminf_{s \to 0} \pi(s) n(s,T).
\end{equation*}
We denote by $\Sigma_\pi$ the class of compact operators $T$ which
fulfil $\Delta_\pi(T) < \infty$.
Analogously, for suitable operator functions $\mathcal{T} : \R_+ \to \mathfrak{S}_\infty$ we define the functionals 
\begin{equation*}
  \BFDelta_\pi(\mathcal{T}) = \limsup_{s \to 0} \pi(s) n(1,\mathcal{T}(s)),\quad
  \BFdelta_\pi(\mathcal{T}) = \liminf_{s \to 0} \pi(s) n(1,\mathcal{T}(s)).
\end{equation*}
We denote by $\mathcal{S}^\pi$ the class of operator functions $\mathcal{T}$
which fulfil $\BFDelta_\pi(\beta \mathcal{T}) < \infty$ for all $\beta >
0$. Furthermore, $\mathcal{S}^\pi_0$ is the class of all $\mathcal{T} \in
\mathcal{S}^\pi$ fulfilling $\BFDelta_\pi(\beta \mathcal{T}) = 0$ for
all $\beta > 0$. We will need the following two lemmata which were
also stated in \cite{LSW} for the polynomial case.
\begin{lemma}\label{lem:spi1}
  For $\mathcal{T} \in \mathcal{S}^\pi$ and $\mathcal{T}_0 \in
  \mathcal{S}^\pi_0$
  \begin{align}
    \lim_{\beta \rightarrow 1} \BFDelta_\pi(\beta \mathcal{T}) =
    \BFDelta_\pi(\mathcal{T})\quad&\text{implies}\quad
    \BFDelta_\pi(\mathcal{T} + \mathcal{T}_0) =
    \BFDelta_\pi(\mathcal{T})\quad\text{and}\label{eq:prop1}\\
    \lim_{\beta \rightarrow 1} \BFdelta_\pi(\beta \mathcal{T}) =
    \BFdelta_\pi(\mathcal{T})\quad&\text{implies}\quad
    \BFdelta_\pi(\mathcal{T} + \mathcal{T}_0) =
    \BFdelta_\pi(\mathcal{T}).
  \end{align}
\end{lemma}
\begin{proof}
  From Ky-Fan's inequality \cite[11.1.3]{BS} it follows 
  \begin{equation*}
    n(1,\mathcal{T}(s)+\mathcal{T}_0(s))
    \leq n(1-\varepsilon,\mathcal{T}(s))+n(\varepsilon,\mathcal{T}_0(s))\quad\text{for}\quad \varepsilon,s > 0
  \end{equation*}
  which leads to
  \begin{align*}
    \BFDelta_\pi(\mathcal{T} + \mathcal{T}_0) \leq \limsup_{s \to 0}
    \big(\pi(s) n(1-\varepsilon,\mathcal{T}(s)) + \pi(s) 
    n(\varepsilon,\mathcal{T}_0(s))\big)
    \leq \BFDelta_\pi({\textstyle \frac{1}{1-\varepsilon}}\mathcal{T}).
  \end{align*}
  Using the condition from \eqref{eq:prop1} we obtain
  \begin{equation*}
    \BFDelta_\pi({\textstyle \frac{1}{1-\varepsilon}} \mathcal{T}) \to
    \BFDelta_\pi(\mathcal{T})\quad\text{as}\quad \varepsilon \to 0
  \end{equation*}
  and therefore 
  \begin{equation*}
    \BFDelta_\pi(\mathcal{T} + \mathcal{T}_0) \leq
    \BFDelta_\pi(\mathcal{T}).
  \end{equation*}
  Analogously it holds
  \begin{equation}
    \BFDelta_\pi(\mathcal{T}) \leq
    \BFDelta_\pi({\textstyle\frac{1}{1-\varepsilon}}(\mathcal{T}+\mathcal{T}_0)) +
    \BFDelta_\pi(-{\textstyle\frac{1}{\varepsilon}}\mathcal{T}_0)
    \to \BFDelta_\pi(\mathcal{T}+\mathcal{T}_0)
  \end{equation}
  which proves the first implication. The second one is established in the same way.
\end{proof}
\begin{lemma}\label{lem:spi2}
  Let $\mathcal{T} : \R_+ \to \mathfrak{S}_\infty$ fulfil
  $\mathcal{T}(s) \to T$ in $\mathfrak{S}_\infty$ as $s \to
  0$. Then for all $\pi: \R_+ \to \R_+$ with $\lim_{s \to 0} \pi(s) = 0$
  it holds $\mathcal{T} \in \mathcal{S}^\pi_0$.
\end{lemma}
\begin{proof}
  Let $\mathcal{U}(s) = T$ for $s > 0$. Obviously, it holds $\mathcal{U} \in \mathcal{S}^\pi_0$. Furthermore,
  \begin{equation*}
    \BFDelta_\pi(\beta(\mathcal{T}-\mathcal{U})) = \limsup_{s
    \to 0} \pi(s) n(\beta^{-1},\mathcal{T}(s)-T) =
    0\quad\text{for}\quad \beta > 0
  \end{equation*}
  because $n(\beta^{-1},\mathcal{T}(s) - T)$ is bounded as $\|\mathcal{T}(s) - T\|
  \to 0$ for $s \to 0$. Therefore, 
  $\mathcal{T}-\mathcal{U} \in \mathcal{S}^\pi_0$.
  Since $\mathcal{U}$ fulfils the condition in \eqref{eq:prop1},
  Lemma \ref{lem:spi1} yields $\mathcal{T} \in \mathcal{S}_0^\pi$.
\end{proof}

\subsection{A modified Birman-Schwinger principle}\label{sec:bsprinc}
Using the results from Subsection \ref{sec:sym} we define a modified 
Birman-Schwinger type operator
\begin{equation}\label{eq:bsoriginal}
  \mathcal{Y}_\alpha(\tau) := \left(\frac{\Lambda-\tau}{\A^{(4)}_0 - \Lambda +
  \tau}\right)^{\frac12}V_\alpha \left(\frac{\Lambda-\tau}{\A^{(4)}_0 -
  \Lambda + \tau}\right)^{\frac12} 
\end{equation}
in $H_4$ for $\tau \in (0,\Lambda)$ and $\alpha \in (0,1)$ where 
\begin{equation}\label{eq:Valpha}
  V_\alpha =  U^\ast \sqrt{\alpha
    f}(I-\alpha\sqrt{f}\,U U^\ast\sqrt{f})^{-1}\sqrt{\alpha f}\,U
\end{equation}
and $U$ is the isometry defined in \eqref{eq:U}.
Since $f \in L^\infty(\R^2;[0,1])$, the operator $V_\alpha$ is
bounded. The following version of the Birman-Schwinger principle \cite{B1,Schw} is
crucial for our considerations. 
\begin{lemma}\label{lem:transpert}
  For $\tau \in (0,\Lambda)$ and $\alpha \in (0,1)$ it holds
  \begin{equation}\label{eq:bsprinc}
    n_-\left(\Lambda - \tau,\A^{(4)}_\alpha\right) = n_+\left(1, \mathcal{Y}_\alpha(\tau)\right).
  \end{equation}
\end{lemma}
\begin{proof}
  For convenience, we put $\A := \A^{(4)}_0$ and $g_\alpha :=
  \sqrt{\alpha f}$. Furthermore, the expression $L \prec H$ means that $L$ is a
  subspace of $H$. 
  Using Glazman's lemma \cite[10.2.2]{BS} we obtain
  \begin{equation*}
    \begin{split}
      n_-(\delta, \A^{(4)}_\alpha) &= \sup_{L \prec D[a_0^{(4)}]} \{\dim L :
      \a^{(4)}_0[u] - \alpha v[u] < \delta\|u\|^2,\quad 0 \neq u \in L
      \}\\
      &= \sup_{L \prec D[a_0^{(4)}]} \{\dim L :
      \|(A-\delta)^\frac12 u\|^2 < \alpha v[u],\quad 0 \neq u \in L\}.
    \end{split}
  \end{equation*}
  Applying the substitution $v := (A-\delta)^\frac12 u$ and Glazman's
  lemma we obtain
  \begin{equation*}
    \begin{split}
      n_-(\delta, \A^{(4)}_\alpha) &= \sup_{L \prec H_4} \{\dim L:\|v\|^2 < \|g_\alpha Q
      (\A-\delta)^{-\frac12} v\|_\times^2, \quad 0 \neq v \in L\}\\
      &=n_+(1,(\A-\delta)^{-\frac12} Q^\ast g_\alpha^2 Q
      (\A-\delta)^{-\frac12})
    \end{split}
  \end{equation*}
  where $\|\cdot\|_\times$ denotes the norm in $L^2(G;\C^{3\times 3})$.
  Next, we define a bounded operator $\mathcal{W} : H_4 \to H_4^\times$ by
  \begin{equation*}
    \mathcal{W} = g_\alpha Q (A-\delta)^{-\frac12} = g_\alpha U A^\frac12
    (A-\delta)^{-\frac12}. 
  \end{equation*} 
  As is well-known \cite[3.10]{BS}, the discrete
  and essential spectrum of $\mathcal{W}^\ast\mathcal{W}$
  and $\mathcal{W}\mathcal{W}^\ast$ coincide, except for the
  point zero. Because 
  \begin{equation*}
    \mathcal{W}\mathcal{W}^\ast = g_\alpha UU^\ast g_\alpha + \delta
    g_\alpha U(A-\delta)^{-1} U^\ast g_\alpha,
  \end{equation*}
  we obtain
  \begin{equation*}
    \begin{split}
      n_-(\delta, \A^{(4)}_\alpha) &=n_+(1, g_\alpha UU^\ast g_\alpha + \delta g_\alpha U
      (\A-\delta)^{-1}U^\ast g_\alpha)\\
      &=\sup_{L \prec H^\times_4} \{\dim L: \|\tilde{V}_\alpha^{-\frac12} w\|^2_\times <
      \|\mathcal{T}_\alpha w\|^2,\, 0 \neq w \in L\}
    \end{split}
  \end{equation*}
  where 
  \begin{equation*}
    \tilde{V}_\alpha = (I-g_\alpha UU^\ast g_\alpha)^{-1}\quad\text{and}\quad
    \mathcal{T}_\alpha = \sqrt{\delta} (A-\delta)^{-\frac12}
    U^\ast g_\alpha.
  \end{equation*}
  Applying the
  substitution $\tilde{w} := \tilde{V}_\alpha^{-\frac12} w$ we get
  \begin{equation*}
    \begin{split}
      n_-(\delta, \A^{(4)}_\alpha) &= \sup_{L \prec H^\times_4}\{\dim L: \|\tilde{w}\|^2_\times <
      \|\mathcal{T}_\alpha \tilde{V}_\alpha^\frac12 \tilde{w}\|^2,\, 0 \neq \tilde{w}
      \in L \}\\
      &= n_+(1,\tilde{V}_\alpha^\frac12
      \mathcal{T}_\alpha^\ast \mathcal{T}_\alpha \tilde{V}_\alpha^\frac12)
    \end{split}
  \end{equation*}
  Since the spectrum of $\tilde{V}_\alpha^\frac12
  \mathcal{T}_\alpha^\ast \mathcal{T}_\alpha
  \tilde{V}_\alpha^\frac12$ and $ \mathcal{T}_\alpha
  \tilde{V}_\alpha \mathcal{T}_\alpha^\ast$ coincides,
  except in the point zero, we obtain
  \begin{equation*}
    n_-(\delta, \A^{(4)}_\alpha) = n_+(1,\mathcal{T}_\alpha
    \tilde{V}_\alpha \mathcal{T}_\alpha^\ast)
  \end{equation*}
  which is \eqref{eq:bsprinc} for $\delta = \Lambda -\tau$.
\end{proof}
We recall that $U$, defined in \eqref{eq:U}, is an isometry
between $H_4$ and $R(U)$. Therefore the operator $U U^\ast$ is an orthogonal
projection onto $R(U)$ fulfilling
\begin{equation*}
  0 \leq UU^\ast \leq I.
\end{equation*}
Here $0$ and $I$ are the zero and identity
operator in $H_4^\times$. Hence, from the definition of
$V_\alpha$ in \eqref{eq:Valpha} we get
\begin{equation}\label{eq:potest}
  U^\ast \alpha f U \leq V_\alpha \leq U^\ast
  {\textstyle \frac{\alpha}{1-\alpha}}f\, U.
\end{equation}
Defining
\begin{equation}\label{eq:bsop}
  \mathcal{Y}(\tau) := \left(\frac{\Lambda-\tau}{\A^{(4)}_0 - \Lambda +
  \tau}\right)^{\frac12}U^\ast f\, U \left(\frac{\Lambda-\tau}{\A^{(4)}_0 -
  \Lambda + \tau}\right)^{\frac12}
\end{equation}
we obtain 
\begin{equation}\label{eq:bsopest}
  \alpha \mathcal{Y}(\tau) \leq \mathcal{Y}_\alpha (\tau) \leq {\textstyle
    \frac{\alpha}{1-\alpha}} \mathcal{Y}(\tau)
\end{equation}
for $\tau \in (0,\Lambda)$ and $\alpha \in (0,1)$. This allows us to
work with the simplified Birman-Schwinger type operator 
$\mathcal{Y}(\tau)$ instead of $\mathcal{Y}_\alpha(\tau)$.

Note that the operator $\mathcal{Y}(\tau)$ is compact. 
This can be obtained by considering the operator
\begin{equation}\label{eq:BSdreh}
  (\Lambda-\tau)\, \sqrt{f} U \left(A_0^{(4)} - \Lambda + \tau\right)^{-1} 
  U^\ast \sqrt{f}
\end{equation}
in $H_4^\times$ which has the same spectrum beyond zero as
$\mathcal{Y}(\tau)$. Since
\begin{equation*}
  U\left(A_0^{(4)} - \Lambda + \tau\right)^{-1} U^\ast = Q
  \left(A_0^{(4)}(A_0^{(4)}-\Lambda+\tau)\right)^{-1} Q^\ast 
\end{equation*}
is an operator of order minus two and $f$ has compact support, the compactness of
\eqref{eq:BSdreh} is a consequence of the Rellich-Kondrachov Theorem
\cite{AF}.
\subsection{Reduction to the spectral minimum of $\A^{(4)}_0$}\label{sec:specmin}
We develop a slightly modified version of \cite{LSW}. As in Subsection \ref{sec:specanal},
$\lambda_1(\xi)$ denotes the lowest branch of eigenvalues of
$\A^{(4)}(\xi)$. The corresponding normed eigenfunction in
$L^2(J;\C^3)$ is $\psi_1(\xi,\cdot)$.  We set
\begin{equation}\label{eq:lambda0}
  \lambda_0(\xi) := \lambda_1(\xi) - \Lambda\quad \text{for}\quad \xi
  \in \Xi
\end{equation}
where $\Xi := \lambda_1^{-1}([\Lambda,\Lambda+\delta))$ for suitable $\delta > 0$. 
The spectral projection of $\A^{(4)}_0$ onto $[\Lambda,\Lambda+\delta)$ is given by
\begin{equation*}
  (\Pi_c u)(x) := \Phi^\ast_{\xi \to \binom{x_1}{x_2}} \left(\chi_\Xi(\xi) \left\langle (\Phi
  u)(\xi,\cdot),\psi_1(\xi,\cdot)\right\rangle_{L^2(J;\C^3)}
  \psi_1(\xi,x_3)\right),\; u \in H_4.
\end{equation*}
Then the unitary operator
\begin{equation*}
  \Pi_0 : \Pi_c H_4 \to
L^2(\Xi),\qquad (\Pi_0 u)(\xi) := \left\langle (\Phi u)(\xi,\cdot),\psi_1(\xi,\cdot)\right\rangle_{L^2(J;\C^3)}
\end{equation*}
maps $\Pi_c H_4$ into the Fourier transformed scalar-valued regime.
Next, we define 
\begin{equation*}
  P(\xi) := \begin{cases} +\sqrt{\lambda_0(\xi)}\quad\text{for}\quad |\xi| \geq \varkappa,\\ 
    -\sqrt{\lambda_0(\xi)}\quad\text{for}\quad |\xi| \leq \varkappa.
  \end{cases}
\end{equation*}
From the properties of $\lambda_1(\xi)$ which were collected in
Subsection \ref{sec:specanal} we obtain $P \in \mathrm{C}^1(\Xi)$. Furthermore,
P depends only on $|\xi|$. Using \eqref{eq:lambda1} and \eqref{eq:lambda0} we obtain
\begin{equation*}
  |\nabla P(\xi)| = q\quad\text{for}\quad |\xi| = \varkappa
\end{equation*}
and therefore $|\nabla P(\xi)| > 0 $ for $\xi \in \Xi$ if we choose
$\delta >0$ sufficiently small. Now we are able to describe
$\A^{(4)}_0$ by a suitable direct integral. Let 
\begin{equation*}
  M_\lambda = \{\xi \in \R^2 : P(\xi) = \lambda\}\quad\text{for}\quad
  \lambda \in \Theta := (-\sqrt{\delta},\sqrt{\delta}).
\end{equation*}
In particular, $M_0 = \{\xi \in \R^2 : |\xi| = \varkappa\}$ is the
set where $\lambda_1(\cdot)$ attains its minimum.
We denote by $\d M_\lambda$ the measure on $M_\lambda$ which is
induced by $\d\xi$ on $\R^2$. Furthermore, we set $\d\mu_\lambda :=
|\nabla P|^{-1}\d M_\lambda$ and 
\begin{equation*}
G(\lambda) := L^2(M_\lambda,\d\mu_\lambda)\quad\text{for}\quad \lambda \in \Theta.
\end{equation*}
The measures $\d\xi$ and $\d\mu_\lambda \d\lambda$ coincide
on $\Xi$. With these definitions we obtain
\begin{equation*}
  \lambda_0 = \int_\Theta^\oplus \lambda^2\,\d \lambda \quad\text{on}\quad L^2(\Xi)
  = \int_\Theta^\oplus G(\lambda)\,\d \lambda
\end{equation*}
where $\lambda^2$ has to be interpreted as the operator of
multiplication by $\lambda^2$ in $G(\lambda)$.
The spaces $G(\lambda)$ are unitarily equivalent. We denote by
$U_\lambda : G(\lambda) \to G(0)$ the corresponding unitary
operator. It follows that the space $L^2(\Xi)$ is equivalent to 
\begin{equation*}
  G_\Theta := L^2(\Theta) \otimes G(0)
\end{equation*}
where the corresponding unitary operator $\Upsilon : L^2(\Xi)
\to G_\Theta$ is given by
\begin{equation*}
  (\Upsilon u)(\lambda) = U_\lambda (u|_{M_\lambda})
  \quad\text{for}\quad u \in L^2(\Xi)\quad\text{and almost every}\quad \lambda \in \Theta.
\end{equation*}
Finally, we define the isometry 
\begin{equation}
F := \Upsilon \Pi_0
\end{equation}
between $\Pi_c H_4$ and $G_\Theta$ which reduces the operator $\A^{(4)}_0 - \Lambda$
to its essential behaviour near the spectral minimum. Namely, if we define
$J_\lambda u := (Fu)(\lambda)$ for $u \in \Pi_c H_4$ and almost every $\lambda \in
\Theta$ it holds
\begin{equation}\label{eq:FA}
  J_\lambda (\A^{(4)}_0 - \Lambda)u = \lambda^2 J_\lambda
  u.
\end{equation}
\subsection{Reduction of $\mathcal{Y}(\tau)$ to the spectral minimum}\label{sec:specminpert}
For $u \in H_4^\times$ it holds
\begin{equation*}
  \begin{split}
    J_\lambda \Pi_c U^\ast \sqrt{f} u &= U_\lambda \left[ \Pi_0 \Pi_c
      (Q (\A^{(4)}_0)^{-\frac12})^\ast \sqrt{f} u\right]_{M_\lambda}\\
    &=U_\lambda\left[\langle \Phi \sqrt{f} u,
      \varphi_1\rangle_{L^2(J;\C^{3\times 3})}\right]_{M_\lambda}
  \end{split}
\end{equation*}
where
\begin{equation}\label{eq:varphi1}
  \begin{split}
    \varphi_1(\xi,x_3) &= \frac{1}{\sqrt{\lambda_1(\xi)}} (\Phi Q \Phi^\ast \psi_1)(\xi,x_3)\\
    &= \frac{1}{\sqrt{2\lambda_1(\xi)}}\left( {\textstyle \binom{i\xi}{\partial_3}}
      \psi_1(\xi,x_3) + \left( {\textstyle \binom{i\xi}{\partial_3}} \psi_1(\xi,x_3)\right)^T\right)
  \end{split}
\end{equation}
for $\xi \in \Xi$ and $x_3 \in J$. Here $\lambda_1(\xi)$ denotes the lowest eigenvalue of
$\A^{(4)}(\xi)$ corresponding to the normed eigenfunction $\psi_1(\xi,\cdot)$ as stated
in Subsection \ref{sec:specanal}.
Because of the compact support of $f$ and the smoothness of $\varphi_1$ in
$\xi$, the function $\langle \Phi \sqrt{f} u, \varphi_1\rangle_{L^2(J;\C^{3\times 3})}$
is also smooth in $\xi$ for every $u \in
H_4^\times$. Therefore, the operator $X : H_4^\times
\rightarrow G(0)$ with
\begin{equation}
  X u := \left[\langle \Phi \sqrt{f} u,\varphi_1\rangle_{L^2(J;\C^{3\times
        3})}\right]_{M_0}\quad\text{for}\quad u \in H_4^\times
\end{equation}
is well-defined. Moreover, there exists a constant $C > 0$ such that 
\begin{equation}\label{eq:XJlambda}
  \|X u - J_\lambda \Pi_c U^\ast \sqrt{f} u\|_{G(0)} \leq
  C\lambda \|u\|_\times\quad\text{for}\quad u \in H_4^\times,\, \lambda \in
  \Theta.
\end{equation}
The essential spectral properties of the Birman-Schwinger operator \eqref{eq:bsop} can be found in the operator
\begin{equation}\label{eq:L}
  \mathcal{L}(\tau) := F^\ast \mathcal{K}(\tau) \mathcal{K}^\ast(\tau) F
  \oplus \mathbb{O} \quad\text{on}\quad \Pi_c H_4 \oplus (I-\Pi_c) H_4,
\end{equation}
where $\mathcal{K}(\tau)$
is given by 
\begin{equation}
  \mathcal{K}(\tau) \; :\; H_4^\times \to G_\Theta
  \; :\; u
  \mapsto \eta_\tau \otimes X u.
\end{equation}
The function
\begin{equation}\label{eq:eta}
  \eta_\tau(\lambda) =
  \left(\frac{\Lambda-\tau}{\lambda^2+\tau}\right)^\frac12
  \quad\text{for}\quad \lambda \in \Theta
\end{equation}
corresponds to the outer terms of the simplified Birman-Schwinger operator
\eqref{eq:bsop}.
We will see that the spectrum of $\mathcal{L}(\tau)$ is basically determined by
the spectrum of the operator
\begin{equation}\label{eq:K}
  K = X X^\ast
\end{equation}
which is a compact integral operator in $L^2(M_0,\d\mu_0)$ with kernel
\begin{equation}
  k(\eta,\xi) = \frac{1}{(2\pi)^2} \int_G f(x)\,
  \mathrm{e}^{i(\eta-\xi)\cdot\binom{x_1}{x_2}} \langle \varphi_1(\eta,x_3),
  \varphi_1(\xi,x_3)\rangle_{\C^{3\times 3}}\,\d x,\quad\eta,\xi \in M_0.
\end{equation}
\section{Statement of the main results}\label{sec:main}
We assume $f \not\equiv 0$ in $L^2$-sense.
Let $(\varkappa_l(\alpha))_l$ be the eigenvalues of $\A^{(4)}_\alpha$ below
$\Lambda$ in nondecreasing order including multiplicities. These eigenvalues are embedded
eigenvalues for the complete operator $\A_\alpha$. Let furthermore
$(\lambda_l(K))_l$ be the eigenvalues of $K$ in non-increasing order including
multiplicities. 
\begin{theorem}\label{th:alpha}
  For all $l \in \N$ it holds that
  \begin{equation}
    \varkappa_l(\alpha) = \Lambda - \alpha^2 (\Lambda\pi \lambda_l(K))^2 +
    o(\alpha^2) \quad\text{as}\quad \alpha \to 0.
  \end{equation}
\end{theorem}
For the calculation of the eigenvalues $\lambda_l(K)$ 
in the case of rotationally symmetric perturbations see Section
\ref{sec:K}. There we express $\lambda_l(K)$ in terms
of a double series and provide estimates on $\lambda_l(K)$ for
sufficiently big $l$. 

For our second result we need to restrict ourselves to perturbations $f$ which
are rotationally symmetric, i.e. $f(x) = f(|x|)$ for $x \in \R^2$. We set 
\begin{equation}
  a := \sup \{|x| : x \in \operatorname{ess} \supp f\}
\end{equation}
and define
\begin{equation}\label{eq:ft}
  f_t := \int_0^1 f(ar)\, r^{t-3}\,\d r\quad\text{for}\quad t \in [3,\infty).
\end{equation}
Here $f$ is interpreted as function $f : \R_+ \rightarrow [0,1]$
depending only on the radial variable. Furthermore, we set
\begin{equation}\label{eq:w}
  w_-(t) := t^4 f_t^2
  \left(\frac{a\varkappa\,\mathrm{e}}{t}\right)^{2t},\qquad w_+(t) := t^8 f_t^2
  \left(\frac{a\varkappa\,\mathrm{e}}{t}\right)^{2t}.
\end{equation}
Note that the inverse functions $w_\pm^{-1}$ exist if we consider
$w_\pm : (t_0,\infty) \to (0,\tau_0)$ for sufficiently big $t_0 > 0$
which corresponds to small $\tau_0 > 0$.
Now we can state the estimate on the counting function for eigenvalues
of $\A^{(4)}_\alpha$. 
\begin{theorem}\label{th:accumul1}
  It holds
  \begin{equation}
    \limsup_{\tau \to 0} \frac{n_-(\Lambda-\tau,\A^{(4)}_\alpha)}{w_+^{-1}(\tau)} =
    1,\qquad \liminf_{\tau \to 0} \frac{n_-(\Lambda-\tau,\A^{(4)}_\alpha)}{w_-^{-1}(\tau)}= 1. 
    \end{equation}
\end{theorem}
From Theorem \ref{th:accumul1} we can derive the following two
estimates on the eigenvalues of $\A^{(4)}_\alpha$.
\begin{corollary}\label{cor:accumul2}
  For every $\varepsilon > 0$ there exists $N_\varepsilon > 0$ such that 
  \begin{equation}\label{eq:accumul2}
    w_-((1+\varepsilon)k) \leq
    \Lambda - \varkappa_k(\alpha) \leq w_+((1-\varepsilon)k)
  \end{equation}
  holds for all $k > N_\varepsilon$.
\end{corollary}
\begin{corollary}\label{cor:accumul3}
  It holds
  \begin{equation}\label{eq:asympformula}
    \ln (\Lambda-\varkappa_k(\alpha)) = -2k\ln k + o(k\ln k)\quad\text{as}\quad k \to \infty.
  \end{equation}
\end{corollary}
\begin{remark}
  Because the estimate in Corollary \ref{cor:accumul3} does not depend
  on $f$, it is valid for all functions $f \in
  L^\infty (\R^2;[0,1])$, $f \not\equiv 0$ in $L^2$-sense, with compact
  support. Every such function can be estimated almost everywhere
  from below and from above by a non-trivial rotationally symmetric function, up to translations.
\end{remark}
\begin{remark}
  It is interesting to note that the estimates in Theorem \ref{th:accumul1},
  Corollary \ref{cor:accumul2} and \ref{cor:accumul3} do not depend on the
  coupling constant $\alpha$. Moreover, the leading term in the asymptotic
  formula \eqref{eq:asympformula} does not at all depend on the perturbation. 
  This is a consequence of the super-polynomial type of eigenvalue
  accumulation, as can be seen in Lemma \ref{lem:varrho} and its applications
  in \eqref{eq:DeltaBS} and \eqref{eq:DeltaKup}.
\end{remark}
\begin{remark}
  The result \eqref{eq:asympformula} is very similar to asymptotics
  obtained in \cite{MR,RW}. There it was shown that the negative eigenvalues $\Lambda_n$ of the
  three-dimensional Pauli operator with constant magnetic field
  perturbed by a compactly supported potential fulfil
  \begin{equation}\label{eq:magnetic}
    \ln (-\Lambda_k) = -2k\ln k + O(k)\quad\text{as}\quad k \to \infty.
  \end{equation}
  Although the Pauli operator is quite different to the elasticity
  operator, there is a common property which may
  explain the similarity of the accumulation rate asymptotics:
  in both cases an operator with strongly degenerated spectral edge is
  perturbed by a compactly supported perturbation. 

  We suppose that differences between the two problems become visible
  by a refined analysis of the asymptotic formulas. In \cite{FP} a detailed investigation of the Pauli
  operator was accomplished yielding also the second term of the
  asymptotic expansion for \eqref{eq:magnetic}. It is an interesting
  open problem whether the methods from \cite{FP} can be used to
  obtain also a refined formula for the elastic problem.
\end{remark}
\section{Proofs of the main estimates}\label{sec:mainproof}

In the following we denote by $\lambda_l(\cdot)$ the eigenvalues of the
corresponding nonnegative compact operator in non-increasing order including
multiplicities.

\subsection{Proof of Theorem \ref{th:alpha}}
From the definition of $\mathcal{L}(\tau)$ in \eqref{eq:L} and from the minimax
principle \cite[9.2.4]{BS} it follows that
\begin{equation}\label{eq:lambdaL}
  \lambda_l(\mathcal{L}(\tau)) = \|\eta_\tau\|^2_{L^2(\Theta)} \lambda_l(K)
  = \frac{2(\Lambda - \tau) \arctan(\sqrt{\delta/\tau})}{\sqrt{\tau}}
  \lambda_l(K)
\end{equation}
and therefore 
\begin{equation*}
  \lambda_l(\sqrt{\tau} \mathcal{L}(\tau)) = \Lambda \pi \lambda_l(K) + o(1)\quad \text{as}\quad \tau \to 0,\quad l \in
  \N.
\end{equation*}
Lemma \ref{lem:remainder2} which we will prove below gives us
\begin{equation*}
  \|\sqrt{\tau}\mathcal{Y}(\tau) - \sqrt{\tau}\mathcal{L}(\tau)\| \leq C
  (\sqrt{\tau} + \sqrt[4]{\tau}) \to 0\quad\text{as}\quad \tau \to 0
\end{equation*}
which leads to
\begin{equation}\label{eq:YlimK}
  \lambda_l(\sqrt{\tau}\mathcal{Y}(\tau)) \to \Lambda \pi
  \lambda_l(K)\quad\text{as}\quad \tau \to 0.
\end{equation}
From the Birman-Schwinger principle \eqref{eq:bsprinc} it follows that
\begin{equation*}
  \lambda_l(\mathcal{Y}_\alpha(\Lambda - \varkappa_l(\alpha))) = 1.
\end{equation*}
With \eqref{eq:bsopest} this yields 
\begin{equation*}
  \lambda_l(\alpha \mathcal{Y}(\Lambda-\varkappa_l(\alpha))) \leq 1 \leq
  {\textstyle \frac{1}{1-\alpha}} \lambda_l(\alpha
  \mathcal{Y}(\Lambda-\varkappa_l(\alpha)))  
\end{equation*}
and therefore
\begin{equation}\label{eq:Ylim1}
  \lambda_l(\alpha \mathcal{Y}(\Lambda-\varkappa_l(\alpha)))\to 1\quad
  \text{as}\quad \alpha \to 0.
\end{equation}
Since $A^{(4)}_\alpha \geq (1-\alpha) A^{(4)}_0$ it holds that $\inf
\sigma(A^{(4)}_\alpha) \geq (1-\alpha)\Lambda$. Hence, 
\begin{equation*}
  \Lambda-\varkappa_l(\alpha) \to 0\quad \text{as}\quad \alpha \to 0.
\end{equation*}
If we replace now $\tau$ by $\Lambda - \varkappa_l(\alpha)$ in
\eqref{eq:YlimK} and compare it with \eqref{eq:Ylim1}, we obtain
\begin{equation*}
  \alpha^{-1} \sqrt{\Lambda-\varkappa_l(\alpha)} \to \Lambda \pi
  \lambda_l(K)\quad\text{as}\quad \alpha \to 0.
\end{equation*}
This concludes the proof. \qed
\subsection{Proof of Corollary \ref{cor:accumul3}}
From Corollary \ref{cor:accumul2} it follows that we can find $N \in \N$
and $\varepsilon_k > 0$ for $k \in \N$ with $\varepsilon_k \to 0$ as $k \to \infty$
such that 
\begin{equation*}
\ln w_-((1+\varepsilon_k)k) \leq \ln (\Lambda - \varkappa_k(\alpha)) \leq \ln
w_+((1-\varepsilon_k)k)
\end{equation*}
holds for all $k > N$. From the definition of $w_\pm$ in \eqref{eq:w} it follows
\begin{equation*}
  \ln w_\pm((1\mp \varepsilon_k)k) = -2k\ln k + 2\ln f_{(1\mp \varepsilon_k)k} +
  o(k\ln k)\quad\text{as}\quad k \to \infty.
\end{equation*}
It remains to show that $\ln f_{(1\mp \varepsilon_k)k} = o(k\ln k)$ as $k \to
\infty$. But because $f \neq 0$ in $L^2$-sense we find some non-empty interval $I =
[at_1,at_2]$, $t_1,t_2 \in [0,1]$ and some constant $c > 0$ such that
$f \geq c\chi_I$ holds almost everywhere. Therefore, we obtain
\begin{equation*}
  \left|\ln f_{(1\mp \varepsilon_k)k}\right| \leq c\left|\ln \int_{t_1}^{t_2} r^{(1\mp
      \varepsilon_k)k-3}\,\d r\right| = o(k\ln k)\quad\text{as}\quad k
      \to \infty.
\end{equation*}
This concludes the proof. \qed
\subsection{Proof of Corollary \ref{cor:accumul2}}
From Theorem \ref{th:accumul1} it follows that for every $\varepsilon > 0$
there exists an $N_\varepsilon  \in \N$ such that 
\begin{equation*}
  \frac{k}{w_+^{-1}(\Lambda - \varkappa_k(\alpha))} = \frac{n_-(\varkappa_k(\alpha)+0,\A^{(4)}_\alpha)}{w_+^{-1}(\Lambda - \varkappa_k(\alpha))}
   \leq \frac{1}{1-\varepsilon }
\end{equation*}
holds for all $k > N_\varepsilon$.
Therefore, the upper estimate in \eqref{eq:accumul2} follows. The lower
estimate is obtained analogously. \qed 

For the proof of Theorem \ref{th:accumul1} we need a small auxiliary result
which states a very special behaviour of the functions
\begin{equation}\label{eq:varrho}
  \varrho_\pm(\tau) := \frac{1}{w_\pm^{-1}(\tau)}.
\end{equation}
\begin{lemma}\label{lem:varrho}
  For all $c > 0$ it holds that
  \begin{equation}
    \lim_{\tau \to 0} \frac{\varrho_\pm(\tau)}{\varrho_\pm(c \tau)} = 1.
  \end{equation}
\end{lemma}
\begin{proof}
  We set 
  \begin{equation*}
    \tilde{w}_\pm(x) := w_\pm({\textstyle \frac{1}{x}}) = \varrho_\pm^{-1}(x)
  \end{equation*}
  and omit ``$\pm$'' in $\varrho_\pm$ and $\tilde{w}_\pm$.
  By the mean value theorem for sufficiently small $\tau > 0$ and certain
  $\xi \in (\min(\tau,c \tau),\max(\tau,c \tau))$ it holds that
  \begin{equation*}
    \left|\ln \frac{\varrho(\tau)}{\varrho(c\tau)}\right| = |\ln
    \varrho(\tau) - \ln \varrho(c \tau)| =
    \frac{|1-c|\tau}{\varrho(\xi) \tilde{w}'(\varrho(\xi))} \leq
    \frac{|1-c^2|\xi}{c \varrho(\xi) \tilde{w}'(\varrho(\xi))}.
  \end{equation*}
  Because $x := \varrho(\xi) \to 0$ as $\tau \to 0$ we obtain
  \begin{equation*}
    0 \leq \lim_{\tau \to 0} \left| \ln
      \frac{\varrho(\tau)}{\varrho(c\tau)}\right| \leq \lim_{x \to 0}
    \frac{|1-c^2|}{c} \frac{\tilde{w}(x)}{x\tilde{w}'(x)} = 0.
  \end{equation*}
  The latter can be derived from the definition of $w_\pm$ in
  \eqref{eq:w} where only the following small difficulty arises. One has to
  verify the boundedness of the expression
  \begin{equation}\label{eq:bounded}
    \left|\frac{\int_0^1 f(ar)r^n \ln r\,\d r}{\int_0^1 f(ar) r^n \,\d
        r}\right| \leq \frac{\int_0^1 f(ar)r^{n-1}\,\d r}{\int_0^1 f(ar) r^n \,\d
        r}
  \end{equation}
  as $n \to \infty$. To obtain this we note that there exists $\varepsilon_0 >
  0$ such that $\int_0^1 f(ar) (r-\varepsilon_0)\,\d r > 0$. This is
  equivalent to
  \begin{equation*}
    \int_0^{\varepsilon_0} f(ar)(\varepsilon_0 - r)\,\d r <
    \int_{\varepsilon_0}^1 f(ar) (r-\varepsilon_0)\,\d r.
  \end{equation*}
  Hence,
  \begin{equation*}
    \begin{split}
      \int_0^{\varepsilon_0} f(ar)r^{n-1}(\varepsilon_0 - r)\,\d r &\leq
      \int_0^{\varepsilon_0} f(ar)\varepsilon_0^{n-1}(\varepsilon_0 -
      r)\,\d r\\ <  \int_{\varepsilon_0}^1 f(ar) \varepsilon_0^{n-1}
      (r-\varepsilon_0)\,\d r &\leq \int_{\varepsilon_0}^1 f(ar) r^{n-1}
      (r-\varepsilon_0)\,\d r
  \end{split}
\end{equation*}
  for all $n \in \N$ which is equivalent to $\int_0^1f(ar) r^{n-1}(r-\varepsilon_0)\,\d r
  > 0$.  This proves the boundedness of \eqref{eq:bounded}.
\end{proof}

\subsection{Proof of Theorem \ref{th:accumul1}}
As we will prove in Section \ref{sec:K} below, the operator $K$ defined in
\eqref{eq:K} fulfils
\begin{equation}\label{eq:DeltaK}
  \BFDelta_{\pi_+}(K) = 1,\qquad \BFdelta_{\pi_-}(K) = 1,\qquad 
  \pi_\pm(\tau) := \varrho_\pm(\tau^2).
\end{equation}
The functionals $\BFDelta$ and $\BFdelta$ were introduced in
Subsection \ref{sec:Sp}. 
From the definition of $\mathcal{L}(\tau)$ in \eqref{eq:L} and from the minimax
principle \cite[9.2.4]{BS} it follows that
\begin{equation*}
  \lambda_l(\mathcal{L}(\tau)) = \|\eta_\tau\|^2_{L^2(\Theta)} \lambda_l(K)
  = \frac{2(\Lambda - \tau) \arctan(\sqrt{\delta/\tau})}{\sqrt{\tau}}
  \lambda_l(K).
\end{equation*}
Therefore, if we set $\zeta_\tau := 2(\Lambda-\tau)\arctan(\sqrt{\delta/\tau})$, we obtain for arbitrary $\beta > 0$
\begin{equation*}
  n(1,\beta \mathcal{L}(\tau)) = n((\beta\zeta_\tau)^{-1}\sqrt{\tau},K).
\end{equation*}
Note that $\zeta_\tau \to \pi\Lambda$ as $\tau \to 0$. Using Lemma \ref{lem:varrho} and \eqref{eq:DeltaK} we obtain
\begin{equation*}
  \begin{split}
    \BFDelta_{\varrho_+}(\beta\mathcal{L}) &= \limsup_{\tau \to 0}
    \varrho_+(\tau) n(1,\beta\mathcal{L}(\tau))\\ &= \limsup_{\tau \to 0}
    \frac{\varrho_+(\tau)}{\varrho_+((\beta\zeta_\tau)^{-2}\tau)}
    \varrho_+([(\beta\zeta_\tau)^{-1}\sqrt{\tau}]^2)
    n((\beta\zeta_\tau)^{-1}\sqrt{\tau},K)\\ &= \BFDelta_{\pi_+}(K) = 1.
  \end{split}
\end{equation*}
In the same way we achieve the result for $\varrho_-$. From the
definition of $\mathcal{L}$ in \eqref{eq:L} it follows that $n(1,\beta
\mathcal{L}(\tau)) = n(1,\sqrt{\beta} F^\ast \mathcal{K}(\tau))$ and therefore
\begin{equation}\label{eq:DeltaFK}
  \BFDelta_{\varrho_+}(\beta F^\ast \mathcal{K}) = 1,\qquad
  \BFdelta_{\varrho_-}(\beta F^\ast \mathcal{K}) = 1\quad\text{for all}\quad
  \beta > 0.
\end{equation}
Let us denote by
\begin{equation*}
 \mathcal{Z}(\tau) := \sqrt{f} U \left(\frac{\Lambda-\tau}{\A^{(4)}_0-\Lambda +\tau}\right)^\frac12
\end{equation*}
the right part of the Birman-Schwinger operator \eqref{eq:bsop}. In Lemma
\ref{lem:remainder1} below we will show that $\mathcal{Z}^\ast(\tau) - F^\ast
\mathcal{K}(\tau)$ converges in $\mathfrak{S}_\infty$ to a compact operator as
$\tau \to 0$. Therefore, we can apply Lemma \ref{lem:spi2} for this operator
family and obtain
\begin{equation*}
  \mathcal{Z}^\ast(\tau) - F^\ast \mathcal{K}(\tau) \in
  \mathcal{S}^{\varrho_+}_0.
\end{equation*}
By \eqref{eq:DeltaFK} we have proven $F^\ast \mathcal{K} \in
\mathcal{S}^{\varrho_+}$ and $\lim_{\beta \to 1} \Delta_{\varrho_+}(\beta
F^\ast \mathcal{K}) = \Delta_{\varrho_+}(F^\ast\mathcal{K})$. Hence, 
Lemma \ref{lem:spi1} is applicable which yields
\begin{equation}
  \BFDelta_{\varrho_+}(\mathcal{Y}) =
  \BFDelta_{\varrho_+}(\mathcal{Z}^\ast) = \BFDelta_{\varrho_+}(
  F^\ast \mathcal{K}) = 1.
\end{equation}
Using Lemma \ref{lem:varrho} and inequality \eqref{eq:bsopest} we achieve
\begin{equation}\label{eq:DeltaBS}
  \BFDelta_{\varrho_+}(\mathcal{Y}) =
  \BFDelta_{\varrho_+}(\alpha \mathcal{Y}) \leq
  \BFDelta_{\varrho_+}(\mathcal{Y}_\alpha) \leq
  \BFDelta_{\varrho_+}({\textstyle \frac{\alpha}{1-\alpha}}
  \mathcal{Y}) = \BFDelta_{\varrho_+}(\mathcal{Y}).
\end{equation}
This and the same procedure for
$\BFdelta_{\varrho_-}(\mathcal{Y}_\alpha)$ yield
\begin{equation*}
  \BFDelta_{\varrho_+}(\mathcal{Y}_\alpha) = 1,\qquad
  \BFdelta_{\varrho_-}(\mathcal{Y}_\alpha) = 1.
\end{equation*}
The proof is completed by applying the Birman-Schwinger principle
\eqref{eq:bsprinc}. \qed

\section{Treatment of the operator $K$}\label{sec:K}

\subsection{Calculation of the eigenvalues}
As mentioned in Subsection \ref{sec:specminpert}, the operator $K$ is an
integral operator in $L^2(M_0,\d\mu_0)$ with kernel 
\begin{equation}
  k(\eta,\xi) = \frac{1}{(2\pi)^2} \int_G f(x)\,
  \mathrm{e}^{i(\eta-\xi)\cdot\binom{x_1}{x_2}} \langle\varphi_1(\eta,x_3) ,
  \varphi_1(\xi,x_3)\rangle_{\C^{3\times 3}}\,\d x,\quad\eta,\xi \in M_0.
\end{equation}
If we remember the definition of $\varphi_1$ in \eqref{eq:varphi1} and of
$\psi_1$ in \eqref{eq:psi1} and set 
\begin{equation*}
  p :=  \varkappa^2\|d_1\|_{J}^2+\|d_2\|_{J}^2,\quad p_0 =
  2\|d_2'\|_{J}^2,\quad p_1 = \|d_1'+d_2\|_{J}^2,\quad p_2 = 
  2\|d_1\|_{J}^2,
\end{equation*}
where $\|\cdot\|_J$ is the norm in $L^2(J)$ and $d_{1/2}(t) :=
d_{1/2}(\varkappa,t)$, we obtain
\begin{equation}
  k(\eta,\xi) = \frac{p_2(\eta\cdot \xi)^2 + p_1 (\eta\cdot \xi) +
    p_0}{\Lambda p (2\pi)^2}\int_{\R^2} f(x) \mathrm{e}^{ix\cdot(\eta-\xi)}
  \,\d x.
\end{equation}
Keeping in mind that $f$ has to be rotationally symmetric and $a = \sup \{|x|
: x \in \operatorname{ess} \supp f\}$ we obtain for the integral on the right hand side
\begin{equation*}
  \begin{split}
    \int_{\R^2} f(x)\mathrm{e}^{ix \cdot (\eta -\xi)}\,\d x &= \int_0^a
    \int_0^{2\pi} f(r) r \mathrm{e}^{ir|\eta - \xi|\cos
      \varphi}\,\d\varphi\,\d r\\
    &=2\pi a^2\sum_{k=0}^\infty
    \frac{(-1)^k}{(k!)^2}\left(\frac{a|\eta-\xi|}{2}\right)^{2k} \int_0^1
    f(ar) r^{2k+1}\,\d r.
  \end{split}
\end{equation*}
With the substitutions
\begin{equation*}
  \eta = \varkappa \binom{\cos s}{\sin s},\quad \xi = \varkappa \binom{\cos
    t}{\sin t},\quad \d\mu_0 = \varkappa |[\nabla P]_{M_0} |^{-1}\,\d t =
  \frac{\varkappa}{q}\,\d t
\end{equation*}
we can consider the operator $K$ also as operator in $L^2((0,2\pi))$. 
Using
\begin{equation*}
  \eta \cdot \xi = \varkappa^2 \cos(s-t),\quad |\eta - \xi|^2 =
  2\varkappa^2(1-\cos(s-t))
\end{equation*}
and setting
\begin{equation*}
  \tilde{f}_k := \int_0^1 f(ar)r^{2k+1}\,\d r,
\end{equation*}
the kernel of $K$ can be rewritten as
\begin{equation}
  k(s,t) = \frac{a^2\varkappa}{2\pi \Lambda p q}\sum_{m=0}^2 p_m
    (\varkappa^2\cos(s-t))^m \sum_{k=0}^\infty
  \frac{(a^2\varkappa^2(\cos(s-t)-1))^k}{2^k(k!)^2} \tilde{f}_k.
\end{equation}
Obviously, $k(s,t)$ is a power series in
$\cos(s-t)$. Therefore, $k(s,t)$ can also be written as
\begin{equation}\label{eq:kpow}
  k(s,t) = \sum_{l=-\infty}^\infty \mu_{|l|}
  \frac{\mathrm{e}^{il(s-t)}}{2\pi}
\end{equation}
which means that $K$ has exactly the eigenfunctions $s \mapsto
\mathrm{e}^{\pm i n s}$ corresponding to the eigenvalues $\mu_n$ for $n
\in \N_0$. The eigenvalue $\mu_n$ is given by
\begin{equation}\label{eq:mu}
  \begin{split}
    &\mu_n = \mu_n \mathrm{e}^{in\cdot 0} = \int_0^{2\pi} k(0,t)
    \mathrm{e}^{int}\,\d t\\
    &=\frac{a^2\varkappa}{\Lambda p q}\sum_{m=0}^2 p_m
    \varkappa^{2m} \sum_{k=0}^\infty
    \frac{(a\varkappa)^{2k}\tilde{f}_k}{2^k(k!)^2}\cdot
    \frac{1}{2\pi}\int_0^{2\pi}(\cos(t))^m(\cos(t)-1)^k\mathrm{e}^{int}\,\d t\\
    &= \frac{a^2\varkappa}{\Lambda p q}  \sum_{m=0}^2 p_m \varkappa^{2m} \sum_{k=|n-m|_+}^\infty
    \frac{(a\varkappa)^{2k}\tilde{f}_k}{2^k(k!)^2} \sum_{l = |\lceil \frac{m-n}{2} \rceil|_+}^{\lfloor
      \frac{k-n+m}{2} \rfloor} {\textstyle\frac{(-1)^{k+n+m}}{2^{2l+n}} \binom{k}{2l+n-m}\binom{2l+n}{l}}.
  \end{split}
\end{equation}
Here $|j|_+ = \frac12(|j| + j)$ for $j \in \N$.

\subsection{Estimates for the eigenvalues}

From a detailed analysis of the latter expression which will be given in
Appendix \ref{app:A} it follows that we can find $c > 0$ and $N \in \N$ such that 
\begin{equation}\label{eq:muestimate}
  c^{-1} (2n)^3f_{2n} \left(\frac{a\varkappa\mathrm{e}}{2n}\right)^{2n} \leq
  \mu_n \leq c (2n)^3 f_{2n}
  \left(\frac{a\varkappa\mathrm{e}}{2n}\right)^{2n}\quad\text{for
  all}\quad n > N.
\end{equation}
Notice that we use the expression $f_{2n}$ in this formula which was defined in \eqref{eq:ft}.
From \eqref{eq:kpow} we see that the eigenspaces of the eigenvalues $\mu_n$ have dimension
two for $n\geq 1$ and dimension one for $n = 0$. If we denote by
$\lambda_n(K)$ for $n \in \N$ the eigenvalues of $K$ in non-increasing order including 
multiplicities, we obtain for a certain constant $c > 0$ and all sufficiently
big $n \in \N$ 
\begin{equation}\label{eq:evestimate}
  c^{-1} n^p f_n \left(\frac{a\varkappa\mathrm{e}}{n}\right)^n \leq
  \lambda_n(K) \leq c n^p f_n
  \left(\frac{a\varkappa\mathrm{e}}{n}\right)^n,\quad p = \begin{cases}3 & \text{if
      $n$ even,}\\4 & \text{if $n$ odd}.\end{cases}
\end{equation}
For odd $n$ this follows from $\lambda_n(K) = \lambda_{n-1}(K)$ and from
application of the estimate for $\lambda_{n-1}(K)$, where we need the fact that
$f_{n-1}/f_n$ is bounded as $n \to \infty$. This was already solved in
the proof of Lemma \ref{lem:varrho}. In the same way we obtain the
existence of $c > 0$ such that for all sufficiently big $n \in \N$ it holds
\begin{equation}\label{eq:evplusestimate}
    c^{-1} n^p f_n \left(\frac{a\varkappa\mathrm{e}}{n}\right)^n \leq
  \lambda_{n+1}(K) \leq c n^p f_n
  \left(\frac{a\varkappa\mathrm{e}}{n}\right)^n,\quad p = \begin{cases}3 & \text{if
      $n$ even,}\\2 & \text{if $n$ odd}.\end{cases}
\end{equation}
\subsection{Estimates for the number of eigenvalues}
For suitable $\tau_0,t_0 > 0$ we define $\pi_\pm : (0,\tau_0) \to
(t_0,\infty)$ as the unique functions fulfilling 
\begin{equation*}
  (\pi_-)^{-1}({\textstyle \frac{1}{t}}) = t^2 f_t
  \left(\frac{a\varkappa\,\mathrm{e}}{t}\right)^t,\quad (\pi_+)^{-1}({\textstyle \frac{1}{t}}) = t^4 f_t
  \left(\frac{a\varkappa\,\mathrm{e}}{t}\right)^t,\quad t \in (t_0,\infty).
\end{equation*}
This coincides with the definitions from \eqref{eq:varrho} and
\eqref{eq:DeltaK}. 

For $\tau \in [\lambda_{k+1}(K),\lambda_k(K)) \cap (0,\tau_0)$ we obtain from \eqref{eq:evestimate} that
\begin{equation*}
  \pi_+(c^{-1}\tau)n(\tau,K) \leq \pi_+(c^{-1}\lambda_k(K)) k \leq 1
\end{equation*}
and therefore, using Lemma \ref{lem:varrho} which is also valid for
$\pi_\pm$,
\begin{equation}\label{eq:DeltaKup}
  \Delta_{\pi_+}(K) = \limsup_{\tau \to 0} \pi_+(\tau) n(\tau,K) = \limsup_{\tau
    \to 0} \pi_+(c^{-1}\tau) n(\tau,K) \leq 1.
\end{equation}
On the other hand, we find arbitrarily small $\tau = \lambda_k(K)-0$, $k$ odd, such that 
\begin{equation*}
  \pi_+(2c\tau) n(\tau,K) = \pi_+(2c(\lambda_k(K)-0)) k \geq 1.
\end{equation*}
Therefore, 
\begin{equation*}
  \Delta_{\pi_+}(K) = \limsup_{\tau \to 0} \pi_+(\tau) n(\tau,K) = \limsup_{\tau
    \to 0} \pi_+(2c \tau) n(\tau,K) \geq 1.
\end{equation*}
This proves the first part of \eqref{eq:DeltaK}. The second part follows by
the analogous investigation for $\delta_{\pi_-}(K)$ using \eqref{eq:evplusestimate}.
\section{Estimates for the remainder terms}\label{sec:remainder}

\begin{lemma}\label{lem:remainder1}
  The operator $\mathcal{Z}^\ast(\tau) - F^\ast \mathcal{K}(\tau)$ converges in
  $\mathfrak{S}_\infty$ as $\tau \to 0$.
\end{lemma}
\begin{proof}
  Let us note that $\mathcal{Z}^\ast(\tau)$ and 
  $F^\ast \mathcal{K}(\tau)$ are compact operators for all $\tau > 0$
  since $\mathcal{Y}(\tau)$ and $\mathcal{L}(\tau)$ are compact.
  We can split $\mathcal{Z}(\tau)$ into 
  \begin{equation*}
    \mathcal{Z}(\tau) = \mathcal{Z}(\tau) \Pi_c + \mathcal{Z}(\tau) (I-\Pi_c).
  \end{equation*}
  From the boundedness of $(\A^{(4)}_0 - \Lambda)^{-\frac12}(I-\Pi_c)$ it
  follows immediately that 
  \begin{equation*}
    \mathcal{Z}(\tau)(I-\Pi_c) = \sqrt{f} U
    \left(\frac{\Lambda-\tau}{\A^{(4)}_0 - \Lambda + \tau}\right)^\frac12(I-\Pi_c)
  \end{equation*}
  converges in $\mathfrak{S}_\infty$ as $\tau \to 0$. The proof is
  accomplished if we can show that also 
  \begin{equation*}
    \mathcal{Q}(\tau) := F\Pi_c \mathcal{Z}^\ast(\tau) - \mathcal{K}(\tau)
  \end{equation*}
  converges in $\mathfrak{S}_\infty$ as $\tau\to 0$. This is done by a
  detailed analysis of the spectral structure developed in Subsections
  \ref{sec:specmin} and \ref{sec:specminpert}.
  We recall that due to \eqref{eq:FA} and the definition of $\eta_\tau$ in
  \eqref{eq:eta} it holds
  \begin{equation*}
    \Pi_c \mathcal{Z}^\ast(\tau) = \Pi_c \left(\frac{\Lambda-\tau}{\A^{(4)}_0
        - \Lambda + \tau}\right)^\frac12 U^\ast \sqrt{f} = F^\ast \eta_\tau F
    \Pi_c U^\ast \sqrt{f}.
  \end{equation*}
  Therefore,
  \begin{equation*}
    \mathcal{Q}(\tau) u = \eta_\tau F \Pi_c U^\ast \sqrt{f}\, u - \eta_\tau
    \otimes X u = \eta_\tau R u,\quad \text{for}\quad u \in H_4^\times,
  \end{equation*}
  where the operator $R : H_4^\times \to G_\Theta$
  is given by 
  \begin{equation*}
    R = F\Pi_c U^\ast \sqrt{f} - I \otimes X.
  \end{equation*}
  For $\tau_1 > \tau_2 > 0$ and $u \in H_4^\times$ we obtain that
  \begin{equation*}
    \begin{split}
      \|(\mathcal{Q}(\tau_1) - \mathcal{Q}(\tau_2))u\|^2_{G_\Theta} &= \int_\Theta
      \|(\eta_{\tau_1}(\lambda) - \eta_{\tau_2}(\lambda)) (R
      u)(\lambda)\|^2_{G(0)}\,\d \lambda\\
      &= \int_\Theta |\eta_{\tau_1}(\lambda) -\eta_{\tau_2}(\lambda)|^2\,\|
      J_\lambda \Pi_c U^\ast \sqrt{f}\,u-X u\|^2_{G(0)}\,\d\lambda\\
      &\leq C\|u\|_\times^2\int_\Theta
      \lambda^2|\eta_{\tau_1}(\lambda)-\eta_{\tau_2}(\lambda)|^2\,\d\lambda
    \end{split}
  \end{equation*}
  for a certain $C > 0$. This is a direct consequence of \eqref{eq:XJlambda}.
  The integral on the right hand side can be estimated by $C_1 \sqrt{\tau}$
  for a certain constant $C_1 > 0$. Therefore, we achieve
  \begin{equation*}
    \|(\mathcal{Q}(\tau_1) - \mathcal{Q}(\tau_2))u\|^2_{G_\Theta} \leq CC_1
    \sqrt{\tau} \|u\|_\times^2
  \end{equation*}
  which proves the convergence of $\mathcal{Q}(\tau)$ as $\tau \to 0$.
\end{proof}
\begin{lemma}\label{lem:remainder2}
  There exists a constant $C > 0$ such that for all sufficiently 
  small $\tau > 0$
  \begin{equation}\label{eq:YminusLmain}
    \|\mathcal{Y}(\tau) - \mathcal{L}(\tau) \| \leq C (1+1/\sqrt[4]{\tau}).
  \end{equation}
\end{lemma}
\begin{proof}
  On $(I-\Pi_c) H_4$ the operator $\mathcal{Y}(\tau)$ is bounded as $\tau \to
  0$ whereas the operator $\mathcal{L}(\tau)$ is identically zero. On the other
  hand, for $u \in \Pi_c H_4$ it holds
  \begin{equation}\label{eq:YminusL}
    \begin{split}
      &|\langle (\mathcal{Y}(\tau) - \mathcal{L}(\tau))u,u\rangle| = |\langle
      \mathcal{Z}(\tau) u, \mathcal{Z}(\tau) u\rangle - \langle
      \mathcal{K}^\ast(\tau) F u, \mathcal{K}^\ast(\tau) F u\rangle|\\
      &\leq \|(\mathcal{Z}(\tau) - \mathcal{K}^\ast(\tau) F)u\|^2 + 2|\mathrm{Re}
      \langle (\mathcal{Z}(\tau)-\mathcal{K}^\ast(\tau)
      F)u,\mathcal{K}^\ast(\tau) F u\rangle|\\
      &\leq \|(\mathcal{Z}(\tau) - \mathcal{K}^\ast(\tau) F)u\|^2 +
      2\|(\mathcal{Z}(\tau) - \mathcal{K}^\ast(\tau) F)u\|\,
      \|\mathcal{K}^\ast(\tau) F u\|.
    \end{split}
  \end{equation}
  From Lemma \ref{lem:remainder1} it follows that $\mathcal{Z}(\tau) -
  \mathcal{K}^\ast(\tau) F$ is bounded as $\tau \to 0$, and from
  \eqref{eq:lambdaL} in the proof of Theorem \ref{th:alpha} it follows that
  \begin{equation*}
    \|\mathcal{K}^\ast(\tau) F\| = \sqrt{\lambda_1(\mathcal{L}(\tau))} \leq
    \frac{c}{\sqrt[4]{\tau}}
  \end{equation*}
  for a certain $c > 0$. Inserting this into \eqref{eq:YminusL} we
  obtain \eqref{eq:YminusLmain}.
\end{proof}

\appendix

\section{}\label{app:A}
We want to derive estimate \eqref{eq:muestimate}. Setting $C_1 :=
a^2\varkappa(\Lambda p q)^{-1}$ we obtain from \eqref{eq:mu} for $n \geq 2$ that
\begin{equation}\label{eq:muapp}
    \mu_n= C_1\sum_{m=0}^2 p_m \varkappa^{2m} \sum_{k=0}^\infty 
    {\textstyle \frac{(-1)^k(a\varkappa)^{2(n+k-m)}\tilde{f}_{n+k-m}}{2^{n+k-m}((n+k-m)!)^2}}
    \sum_{l=0}^{\lfloor \frac{k}{2} \rfloor} {\textstyle \frac{1}{2^{n+2l}}
      \binom{k+n-m}{2l+n-m}\binom{2l+n}{l}}.
\end{equation}
With the estimate
\begin{equation}\label{eq:estapp}
\begin{split}
  \sum_{l=0}^{\lfloor \frac{k}{2} \rfloor} {\textstyle \frac{1}{2^{n+2l}}
    \binom{k+n-m}{2l+n-m}\binom{2l+n}{l}} &\leq
  \frac{(k+n-m)!\,(k+n)(k+n-1)}{2^n\, n!}\sum_{l=0}^{\lfloor \frac{k}{2}
    \rfloor} \frac{1}{4^l l!}\\
  &\leq \frac{(k+n-m)!\,n^2(k+1)^2}{2^n\,n!}\,\mathrm{e}^{\frac14}
\end{split}
\end{equation}
and with the definition $C_2 := \sum_{m=0}^2 p_m 2^m a^{-2m}$ we obtain
\begin{equation*}
  \mu_n\leq C_1C_2\mathrm{e}^\frac14 \left(\frac{a\varkappa}{2}\right)^{2n}
  \frac{n^2 \tilde{f}_{n-2}}{n!} \sum_{k=0}^\infty
  \frac{(a\varkappa)^{2k} (k+1)^2}{2^k (n+k-2)!}.
\end{equation*}
Using $(n+k-2)! \geq (n-2)!\,k!$ and Stirling's formula
\begin{equation*}
  \sqrt{2\pi n}\left(\frac{n}{\mathrm{e}}\right)^n \leq 
  n!\leq \mathrm{e}^{\frac{1}{12 n}}\sqrt{2\pi n} 
  \left(\frac{n}{\mathrm{e}}\right)^n
\end{equation*}
we achieve
\begin{equation*}
  \begin{split}
    \mu_n &\leq C_1C_2\, 4\mathrm{e}^{\frac14 + (a\varkappa)^2} \left(\frac{a\varkappa}{2}\right)^{2n}
    \frac{n^4 \tilde{f}_{n-2}}{(n!)^2}\\
    &\leq C_1C_2 \frac{\mathrm{e}^{\frac14+(a\varkappa)^2}}{4\pi} \, (2n)^3
    \tilde{f}_{n-2} \left(\frac{a\varkappa\,\mathrm{e}}{2n}\right)^{2n}.
  \end{split}
\end{equation*}
This proves the upper estimate in \eqref{eq:muestimate}. In order to prove
the lower estimate, we start again with formula \eqref{eq:muapp} and estimate
it from below. If we define $C_3 := p_2 \varkappa^{4}(a\varkappa)^{-4}$ and use
\eqref{eq:estapp}, we obtain
\begin{equation*}
    \mu_n\geq C_1C_3\left(\frac{a\varkappa}{2}\right)^{2n} \frac{n^4\tilde{f}_{n-2}}{(n!)^2}
     - C_1 C_2 \mathrm{e}^{\frac14}\left(\frac{a\varkappa}{2}\right)^{2n}
     \frac{n^2 \tilde{f}_{n-2}}{n!} \sum_{k=1}^\infty \frac{(a\varkappa)^{2k}(k+1)^2}{2^k(n+k-2)!}.
\end{equation*}
Estimating the sum in the second term and using Stirling's formula, we achieve
\begin{equation*}
\begin{split}
    \mu_n &\geq \left(C_1C_3-\frac{1}{n}C_1C_2\,4
    (a\varkappa)^2\mathrm{e}^{\frac14 +(a\varkappa)^2}\right)\frac{n^4
    \tilde{f}_{n-2}}{(n!)^2}\left(\frac{a\varkappa}{2}\right)^{2n} \\
    &\geq \frac{C_1C_3-\frac{1}{n}C_1C_2\,4(a\varkappa)^2\mathrm{e}^{\frac14+(a\varkappa)^2}}{16\pi
    \mathrm{e}^{\frac{1}{6n}}} \, (2n)^3 \tilde{f}_{n-2} \left(\frac{a\varkappa\,\mathrm{e}}{2n}\right)^{2n}.
  \end{split}
\end{equation*}
This proves the lower bound in \eqref{eq:muestimate} for sufficiently large
$n \in \N$.

\address{
  Institute for Analysis, Dynamics and Modelling\\Department of Mathematics and Physics\\University Stuttgart\\
  Pfaffenwaldring 57\\70569 Stuttgart, Germany\\ email:{Foerster@mathematik.uni-stuttgart.de}}
\end{document}